\documentclass[proof]{WileyASNA-v1}

\articletype{Article Type}%

\received{11 January 2024}
\revised{1 March 2024}
\accepted{5 March 2024}

\defcitealias{2019MNRAS.484.4046M}{Paper\,II}                                   
\defcitealias{2019MNRAS.488.3857B}{Paper\,III}
\defcitealias{2023MNRAS.518.3722B}{Paper\,IV}

\begin{document}

\title{The \textit{HST} Large Programme on NGC\,6752 - V.\\ Differences in Luminosity and Mass Functions among Multiple Stellar Populations}

\author[1,2]{M. Scalco*}
\author[3,4]{R. Gerasimov}
\author[2]{L. R. Bedin}
\author[5]{E. Vesperini}
\author[6,2]{D. Nardiello}
\author[7]{M. Salaris}
\author[3]{\\ A. Burgasser}
\author[8]{J. Anderson}
\author[2,9]{M. Libralato}
\author[8]{A. Bellini}
\author[1]{P. Rosati}

\authormark{SCALCO \textsc{et al}}

\address[1]{Dipartimento di Fisica e Scienze della Terra, Università di Ferrara, Via Giuseppe Saragat 1, Ferrara I-44122, Italy}
\address[2]{Istituto Nazionale di Astrofisica, Osservatorio Astronomico di Padova, Vicolo dell’Osservatorio 5, Padova I-35122, Italy}
\address[3]{Department of Astronomy \& Astrophysics, University of California, San Diego, La Jolla, California 92093, USA}
\address[4]{Department of Physics and Astronomy, University of Notre Dame, Nieuwland Science Hall, Notre Dame, IN 46556, USA}
\address[5]{Department of Astronomy, Indiana University, Swain West, 727 E. 3rd Street, Bloomington, IN 47405, USA}
\address[6]{Dipartimento di Fisica e Astronomia "Galileo Galilei", Universit{\`a} di Padova, Vicolo dellOsservatorio 3, I-35122, Padova, Italy}
\address[7]{Astrophysics Research Institute, Liverpool John Moores University, 146 Brownlow Hill, Liverpool L3 5RF, UK}
\address[8]{Space Telescope Science Institute, 3700 San Martin Drive, Baltimore, MD 21218, USA}
\address[9]{AURA for the European Space Agency (ESA), Space Telescope Science Institute, 3700 San Martin Drive, Baltimore, MD 21218, USA}

\corres{*\email{michele.scalco@inaf.it}}


\abstract{We exploit the astro-photometric dataset of the multi-epoch infrared parallel field of a \textit{Hubble Space Telescope} Large Programme aimed at studying the faintest stars of the globular cluster NGC\,6752 to determine the luminosity and mass functions of the multiple stellar populations of this cluster. Thanks to the measurement of proper motions and deeper completeness, the results presented in this paper represent a significant improvement over those of previous studies. We successfully derived membership probabilities reaching stars as faint as $m_{\rm F160W} \sim 25$, allowing us to reliably distinguish the three main stellar populations detected within this cluster. We employed a new set of model isochrones that have been individually fit to the colour-magnitude diagram of each population. We present a comprehensive analysis of the luminosity and mass functions for three stellar populations within NGC\,6752. Notably, our findings reveal differences in the present-day luminosity and mass functions of first-generation and second-generation stars; these differences are consistent with the manifestation of the effects of dynamical processes acting on populations with different initial spatial distributions. Finally, we publicly release the catalogues with positions, photometry, proper motions, and memberships probabilities, as well as the stacked-image atlases and all newly calculated stellar models.}


\fundingInfo{
This work is based on funding by MIUR under PRIN-2017 
programme $\#$2017Z2HSMF and by INAF under PRIN-2019 programme $\#$10-Bedin.
%
}

\maketitle

\section{Introduction}\label{Section1}

Globular-cluster (GC) stars are commonly categorised into two distinct main groups based on their chemical composition. The first-population (1P) stars exhibit a Galactic-field-like composition, while the second-population (2P) stars are characterised by depletion in specific light elements, such as carbon, oxygen, and magnesium, and enrichment in helium, nitrogen, aluminium, and sodium compared to the 1P stars. Both 1P and 2P stars can host sub-populations of stars \citep[see reviews by][]{1987PASP...99...67S,2018ARA&A..56...83B,2019A&ARv..27....8G}.

Various formation scenarios have been proposed to explain the origin of multiple stellar populations (mPOPs) in GCs. These scenarios can be divided into two categories. The first category of scenarios involves multiple star-formation episodes, where intermediate- to high-mass stars formed during the first burst eject winds of processed material, contributing to the formation of the 2P stars. Gas from massive interacting binaries \citep{2009A&A...507L...1D,2022MNRAS.513.2111R,2023MNRAS.525L.117R}, fast-rotating massive stars \citep{2007A&A...464.1029D,2013A&A...552A.121K}, stellar mergers \citep{2020MNRAS.491..440W}, and asymptotic-giant-branch (AGB) and super-AGB stars (e.g. \citealt{2001ApJ...550L..65V,2008MNRAS.391..825D,2016MNRAS.458.2122D}) have been suggested as possible contributors \citep[see also][for a review]{2015MNRAS.454.4197R}.

The second category of formation scenarios proposes the early accretion of material ejected by supermassive stars or massive interacting binaries by a fraction of stars of the same generation, leading to the formation of all stars in a single star-formation episode (e.g. \citealt{2013MNRAS.436.2398B,2014MNRAS.437L..21D,2018MNRAS.478.2461G,2023MNRAS.521.1646W}).

In the past, the study of mPOPs in GCs focused mainly on stars with masses greater than $\sim$0.6\,M$_{\odot}$, utilising ultraviolet (UV) and visual filters, mainly because of the limited signal-to-noise ratio of the UV observations for stars with lower masses. However, the identification and separation of mPOPs at lower-mass ranges, along with the investigation of their chemical composition, is of fundamental importance to understanding the formation and dynamical history of GCs. Specifically, extending determinations of their mass function (MF) to lower masses and determining the MF slopes of mPOPs across an extensive mass range (from $\sim$0.1 to $\sim$0.8 solar masses) can offer invaluable insights to allow discrimination among various proposed formation scenarios. As outlined in \citet{2018MNRAS.476.2731V}, utilizing N-body simulations, substantial distinctions can arise between the local MFs of the 1P and the 2P when assuming both populations form with the same initial MF. Specifically, within the outer regions of a cluster, the local MF of the 2P tends to exhibit a steeper slope compared to the 1P MF. This phenomenon is a result of the combined effects of mass segregation and the outward migration of low-mass stars, within a system where the 2P was initially more centrally concentrated and populated the outer regions preferentially with low-mass stars.

Very low-mass (VLM) stars are characterised by high-density and low-effective temperature atmospheres. Their spectral peak occurs in the near-infrared (NIR) region, where several oxygen-containing molecules (e.g. CO, H$_2$O, OH, TiO, VO, ZrO), significantly impact opacity \citep{1995ApJ...445..433A}. Consequently, variations in chemical composition are easy to see in the spectra of these stars.

The recent breakthrough in studying mPOPs at lower masses has been made possible by the NIR channel of the Wide Field Camera 3 (WFC3/NIR) on board the \textit{Hubble Space Telescope} (\textit{HST}, \citealt{2012ApJ...754L..34M,2014MNRAS.439.1588M,2017MNRAS.469..800M,2019MNRAS.484.4046M}). The WFC3/NIR F160W band is significantly affected by absorption from various oxygen-containing molecules, including H$_2$O and CO, while WFC3/NIR F110W photometry remains mostly unaffected by the oxygen abundance. As a result, 2P stars, which are depleted in oxygen compared to 1P stars, exhibit brighter F160W magnitudes and redder F110W-F160W colours than the 1P stars. Observations in these filters have proven highly effective in distinguishing and characterising M-dwarf mPOPs in several GCs, including NGC\,2808 \citep{2012ApJ...754L..34M,2022ApJ...927..207D}, NGC\,6121 (M4, \citealt{2014MNRAS.439.1588M,2022ApJ...927..207D}), $\omega$\,Centauri \citep{2017MNRAS.469..800M,roman_omega_cen}, NGC\,6752 \citep{2019MNRAS.484.4046M}, and 47~Tucanae \citep{roman_theory_paper}. Recent \textit{James Webb Space Telescope} (\textit{JWST}) observations have further enriched the observational characterization of multiple populations by revealing distinct populations in M-dwarf stars of a few clusters (see e.g. \citealt{2022MNRAS.517..484N,2023MNRAS.525.2585N,2023MNRAS.522.2429M,2023ApJ...953...62Z,2023arXiv231013056C}).

Here we present the reduction and analysis of NIR images from \textit{HST} observations of a deep field of the nearby ($\sim$4\,kpc) GC NGC\,6752. The data presented here were collected under the \textit{HST Large Programme on NGC\,6752} (GO-15096 + GO-15491; P.I.:\,L. R. Bedin). 

NGC\,6752 is known to host three different stellar populations (A, B, and C), which are clearly identifiable along an extensive part of the CMD, from the red-giant branch (RGB) to the low main sequence (MS). Population A is traditional 1P, with abundances similar to those of field stars with the same metallicity. Populations B and C are the 2Ps and are characterized by enhanced abundances of helium, nitrogen, sodium, and a depletion of carbon, and oxygen; the enhancement/depletion is stronger in population C and milder in population B (see \citealt{2013ApJ...767..120M}).

The primary data set of the programme includes observations of a primary field obtained with the Wide Field Channel (WFC) of the Advanced Camera for Surveys (ACS) with the aim of examining the white dwarf (WD) cooling sequence (CS) within NGC\,6752.  This investigation has been addressed in two earlier papers from this series (\citealt[hereafter \citetalias{2019MNRAS.488.3857B}]{2019MNRAS.488.3857B}, and \citealt[hereafter \citetalias{2023MNRAS.518.3722B}]{2023MNRAS.518.3722B}). To effectively remove background and foreground objects from the observed fields, these programmes were devised to acquire observations at distinct epochs. For each epoch, images of a parallel field were taken with the WFC3/NIR. This parallel field was collected to investigate the mPOPs of NGC\,6752 at the bottom of the MS.

The first epoch of the parallel NIR field was analysed in the second publication of this series \citep[hereafter \citetalias{2019MNRAS.484.4046M}]{2019MNRAS.484.4046M}. In that study, it was found that the three main populations (the aforementioned A, B, and C) --- previously observed in the brightest part of the CMD \citep{2013ApJ...767..120M} --- define three distinct sequences in the IR CMD. These sequences extend from the MS knee, a typical feature found in the NIR CMD and occurring at the lower mass end of the MS. The MS knee is a result of absorption processes involving molecular hydrogen within the atmospheres of cooler, low-mass MS stars, leading to a shift of these stars towards bluer colours \citep{1969ApJ...156..989L,1976ApJ...208..399M}. The distinct sequences extend all the way to the bottom of the MS ($\sim$0.1\,M$_{\odot}$).

In this analysis, we provide a more comprehensive reduction and analysis of the parallel NIR field discussed in \citetalias{2019MNRAS.484.4046M}, utilising the complete data set gathered through the combined GO-15096 + GO-15491 programmes. The inclusion of additional observations has led to significant improvements upon \citetalias{2019MNRAS.484.4046M}. With a temporal baseline of approximately three years, we can now accurately distinguish between cluster members and background/foreground sources using proper motion (PM) measurements. Additionally, the number of available images has doubled since \citetalias{2019MNRAS.484.4046M}. This expanded photometric data set enables us to further distinguish and analyse the three distinct stellar populations with greater precision.

The paper is organised as follows: the observations and data reduction are presented in Section\,\ref{Section2}. In Section\,\ref{Section3} we present the NIR CMD, while Section\,\ref{Section4} and Section\,\ref{Section5} present the results of our determination of the luminosity functions (LFs) and MFs for the three different mPOPs. Finally, Section\,\ref{Section6} provides a brief summary and discussion of our results.

\section{Observations and data reduction}\label{Section2}

A total of 255 images were acquired in three different epochs for a total duration of 85\,\textit{HST} orbits, specifically in 2018 (40\,orbits), 2019 (5\,orbits), and 2021 (40\,orbits).  During the first and third epochs, data were collected using two filters, F110W and F160W.  Only the F160W filter was used during the second epoch.  During each orbit, \textit{HST} collected one short exposure of 143\,s and two long exposures of 1303\,s each. All images were acquired in MULTIACCUM mode with SAMPSEQ=SPARS10 and NSAMP=15 for the short exposures and SAMPSEQ=SPARS100 and NSAMP=14 for the long exposures. Table\,\ref{Table1} reports the complete list of \textit{HST} WFC3/NIR observations for the observed outer field of NGC\,6752.
\begin{table}
\caption{List of \textit{HST} observations of NGC\,6752.}
\label{Table1}
\begin{tabular*}{.48\textwidth}{@{\extracolsep{\fill}}l c c}
 Filter & Exposures & Epoch\\
 \hline
 \hline
 & WFC3/NIR &\\
 \hline
 Epoch\,1: 2018.7 & (GO-15096) & (40\,\textit{HST} orbits)\\
 \hline
 F110W & 28$\times$143 + 56$\times$1303\,s & 2018/09/09-18\\
 F160W & 12$\times$143 + 24$\times$1303\,s & 2018/09/07-09\\
 \hline
 Epoch\,2: 2019.6 & (GO-15096) & (5\,\textit{HST} orbits)\\
 \hline
 F160W & 5$\times$143 + 10$\times$1303\,s & 2019/08/01-16\\
 \hline
 Epoch\,3: 2021.7 & (GO-15491) & (40\,\textit{HST} orbits)\\
 \hline
 F110W & 28$\times$143 + 56$\times$1303\,s & 2021/09/02-11\\
 F160W & 12$\times$143 + 24$\times$1303\,s & 2021/09/02-12\\
 \hline
 \end{tabular*}
\end{table}

\begin{figure}
\centering
\includegraphics[width=\columnwidth]{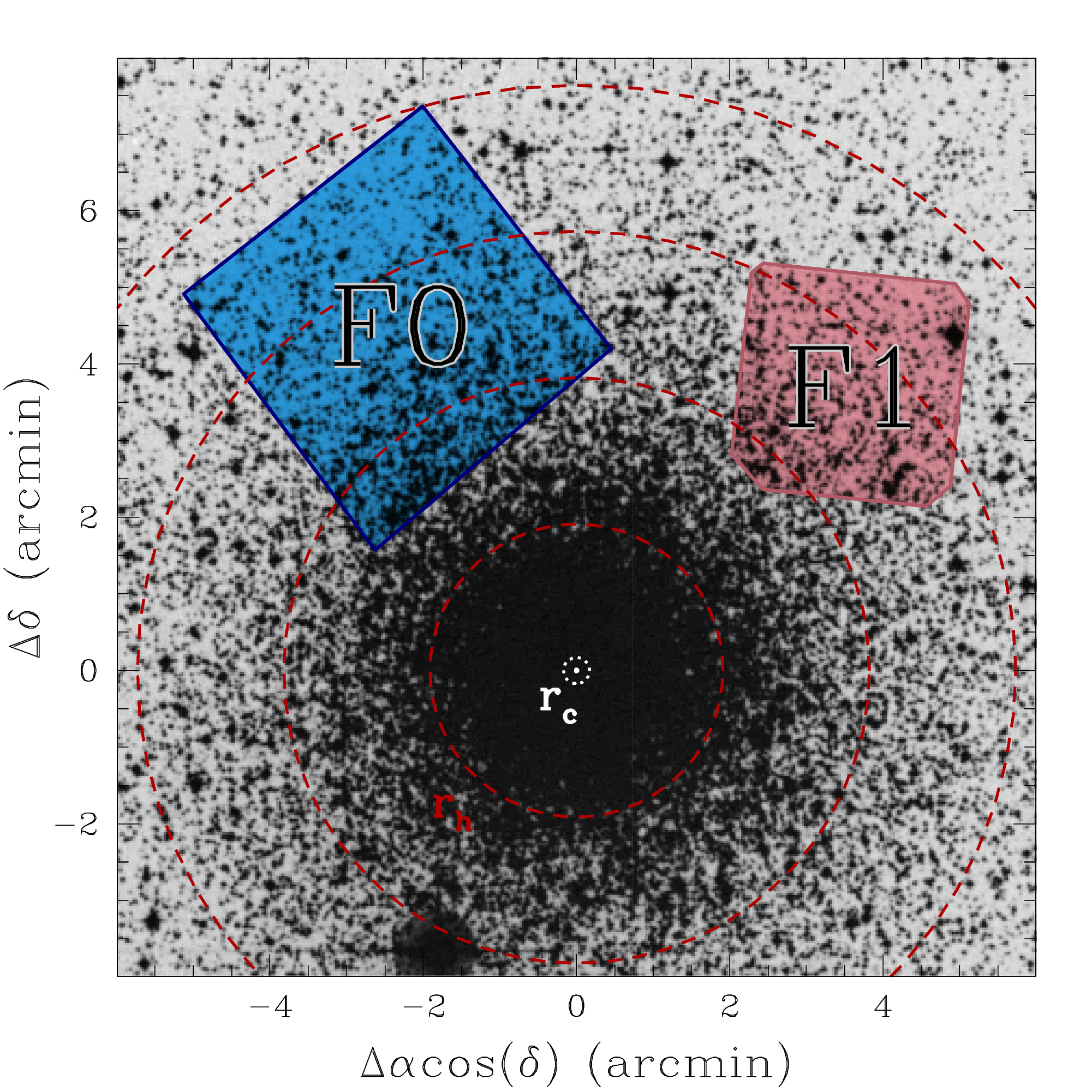}
\caption{Outlines of the fields observed in \textit{HST} programme GO-15096 + GO-15491, superimposed on a DSS image of NGC\,6752. The primary ACS/WFC field (F0) is in azure, while the parallel WFC3/NIR field is shown in pink (F1). Units are in arcmin measured from the cluster centre. The data discussed in this paper come from field F1. The white dashed circle marks the cluster’s core radius ($r_{\rm c}=0^\prime_{\cdot}17$; \citealt{1996AJ....112.1487H,2010arXiv1012.3224H}); the red dashed circles mark the half-light radius ($r_{\rm h}=1^\prime_{\cdot}91$; \citealt{1996AJ....112.1487H,2010arXiv1012.3224H}), and $2\,r_{\rm h}$, $3\,r_{\rm h}$ and $4\,r_{\rm h}$ from the centre.} 
\label{FOV} 
\end{figure}

Figure\,\ref{FOV} shows the locations of the primary and parallel fields, superposed on an image from the Digital Sky Survey (DSS)\footnote{\href{https://archive.eso.org/dss/dss}{https://archive.eso.org/dss/dss}}. The primary ACS/WFC field (F0) is shown in azure, while the parallel WFC3/NIR field (F1) is plotted in pink. In this article, we consider only data from field F1.

The data were reduced following the procedure outlined in \citet{2021MNRAS.505.3549S}. In summary, this procedure involves two main steps: the \textit{first-pass} and \textit{second-pass} photometry. During the \textit{first-pass} photometry, we perturbed a set of``libary'' WFC3/IR effective Point Spread Functions (ePSFs) (see \citealt{2000PASP..112.1360A,2006acs..rept....1A} and \citealt{2016wfc..rept...12A}) to determine the optimal spatially variable PSF for each image. Then, using these PSFs, we extracted the positions and fluxes of the stars within each image. This extraction was carried out using the \texttt{FORTRAN} code \texttt{hst1pass} (see Anderson 2022 WFC3/ISR 2022-05). To account for geometric distortion, the stellar positions in each individual exposure catalogue were corrected using the publicly available WFC3/IR correction \citep{2016wfc..rept...12A}. For each of the two filters, the positions and magnitudes were transformed to a common reference frame using six-parameter linear transformations and photometric zero points.

Subsequently, we performed the \textit{second-pass} photometry using the \texttt{FORTRAN} software package \texttt{KS2}, which is based on the software \texttt{kitchen\_sync} presented in \citealt{2008AJ....135.2055A}.  This software routine makes use of the results obtained from the \textit{first-pass} stage to simultaneously identify and measure stars across all individual exposures and filters. By relying on multiple exposures, \texttt{KS2} effectively detects and measures faint stars that would be otherwise lost in the noise of individual exposures. The star-finding process is executed through a series of passes, gradually moving from the brightest to the faintest stars. In each iteration, the routine identifies stars that are fainter than those found in the previous iteration, subsequently measuring and subtracting them. This iterative approach ensures that progressively fainter stars are detected and accounted for, enhancing the overall accuracy of the photometric measurements. \texttt{KS2} employs three distinct methods for measuring stars, with each approach specifically tailored for different magnitude ranges. We refer to \citet{2017ApJ...842....6B,2018MNRAS.481.3382N,2021MNRAS.505.3549S} for a detailed description of the methods and procedures.

The photometry has been zero-pointed into the Vega magnitude system by following the recipe of \citet{2005MNRAS.357.1038B} and adopting the photometric zero-points provided by STScI web page for WFC3/NIR\footnote{\href{https://www.stsci.edu/hst/instrumentation/wfc3/data-analysis/photometric-calibration}{https://www.stsci.edu/hst/instrumentation/wfc3/data-analysis/photometric-calibration}}. We cross-referenced the stars in our catalogue with the stars in the \textit{Gaia} Data Release 3 (\textit{Gaia} DR3, \citealt{2016A&A...595A...2G,2023A&A...674A...1G}). The sources found in common were used to anchor our positions (X, Y) to the \textit{Gaia} DR3 absolute astrometric system.

To ensure a well-measured sample of stars measured in \textit{HST} images, we implemented a selection process using a set of quality parameters provided by \texttt{KS2}, following a similar approach as described in \citet{2021MNRAS.505.3549S}. The quality parameters employed include the photometric error ($\sigma$), the quality-of-fit (QFIT) parameter, which quantifies the PSF-fitting residuals, and the RADXS parameter, a shape parameter that allows for differentiation between stellar sources, galactic sources, and cosmic ray/hot pixels introduced in \citet{2008ApJ...678.1279B}.  Further details regarding these parameters can be found in \citet{2017ApJ...842....6B,2018MNRAS.481.3382N,2021MNRAS.505.3549S}. 

\begin{figure*}
 \centerline{\includegraphics[width=\textwidth]{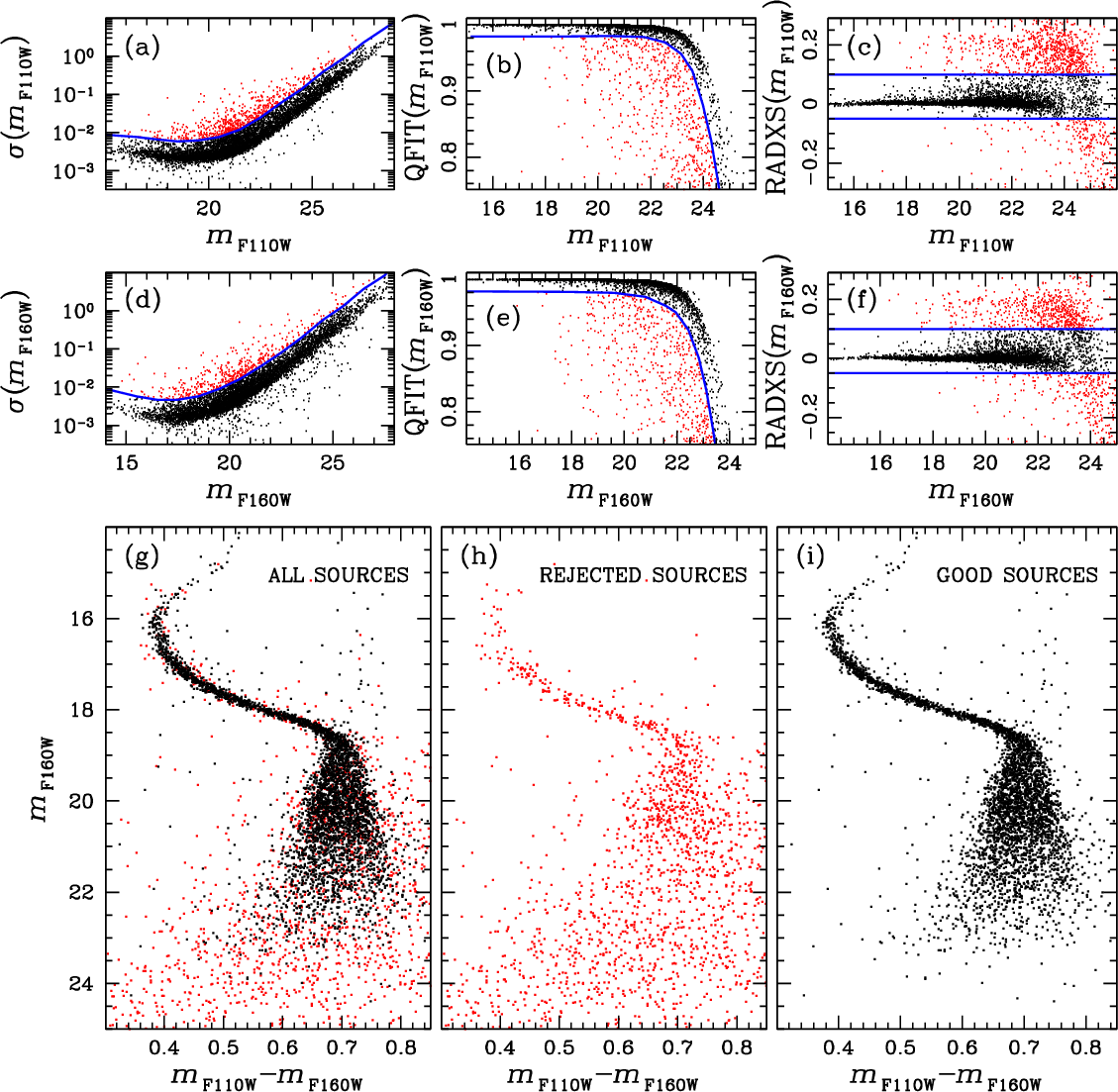}}
 \caption{Procedure used to distinguish well-measured stars from outliers. (a)-(c) distributions of the photometric error ($\sigma$), quality-of-fit (QFIT), and RADXS as a function of the F110W magnitude. The blue lines separate well-measured stars from outliers. (d)-(f) same as (a)-(c) but for the F160W filter. (g)-(i) $m_{\rm F160W}$ versus $m_{\rm F110W}-m_{\rm F160W}$ CMD for all the detected sources, the sources rejected by the selections, and the sources that passed the selections, respectively. In all panels, well-measured sources are represented in black, while rejected sources are shown in red.} 
 \label{Figure1} 
\end{figure*}

Panels\,(a)-(f) of Fig.\,\ref{Figure1} illustrate the selection process for the photometry derived from the short exposures using the first method of \texttt{KS2}. We plotted each parameter as a function of the stellar magnitude. For the $\sigma$ and QFIT parameters, we drew by hand fiducial lines to separate the bulk of well-measured stars from the outliers (panels\,(a), (b), (d) and (e) of Fig.\,\ref{Figure1}). For the RADXS parameter, we selected stars that satisfy the condition: $-$0.05 < RADXS < $+$0.1 (panels\,(d) and (f) of Fig\,\ref{Figure1}). Finally, panels\,(g), (h) and (i) of Fig.\,\ref{Figure1} show the $m_{\rm F160W}$ versus $m_{\rm F110W}-m_{\rm F160W}$ CMD for all the detected sources, the sources rejected by the selection criteria, and the sources that passed the criteria, respectively. We performed the same procedure for the photometry obtained from the long exposures and for all the extraction methods utilised by \texttt{KS2}. We combined the most accurately measured stars from the three different photometric methods of \texttt{KS2} to obtain a single final catalogue, incorporating photometry obtained from both long and short exposures. 

PMs were computed using the technique described in \citet{2021MNRAS.505.3549S} (see also \citealt{2014ApJ...797..115B,2018ApJ...853...86B,2018ApJ...854...45L,2022ApJ...934..150L}). This iterative procedure treats each image as an independent epoch and can be summarised in two main steps: first, it transforms the stellar positions from each exposure into a common reference frame through a six-parameter linear transformation. Then, it fits these transformed positions as a function of the epoch using a least-square straight line. The slope of this line, determined after multiple outlier-rejection stages, provides a direct measurement of the PM. High-frequency-variation systematic effects were corrected as described in \citet{2018ApJ...853...86B}, i.e. according to the median value of the closest 100 likely cluster members (excluding the target star itself).

\begin{figure}
\centering
 \includegraphics[width=\columnwidth]{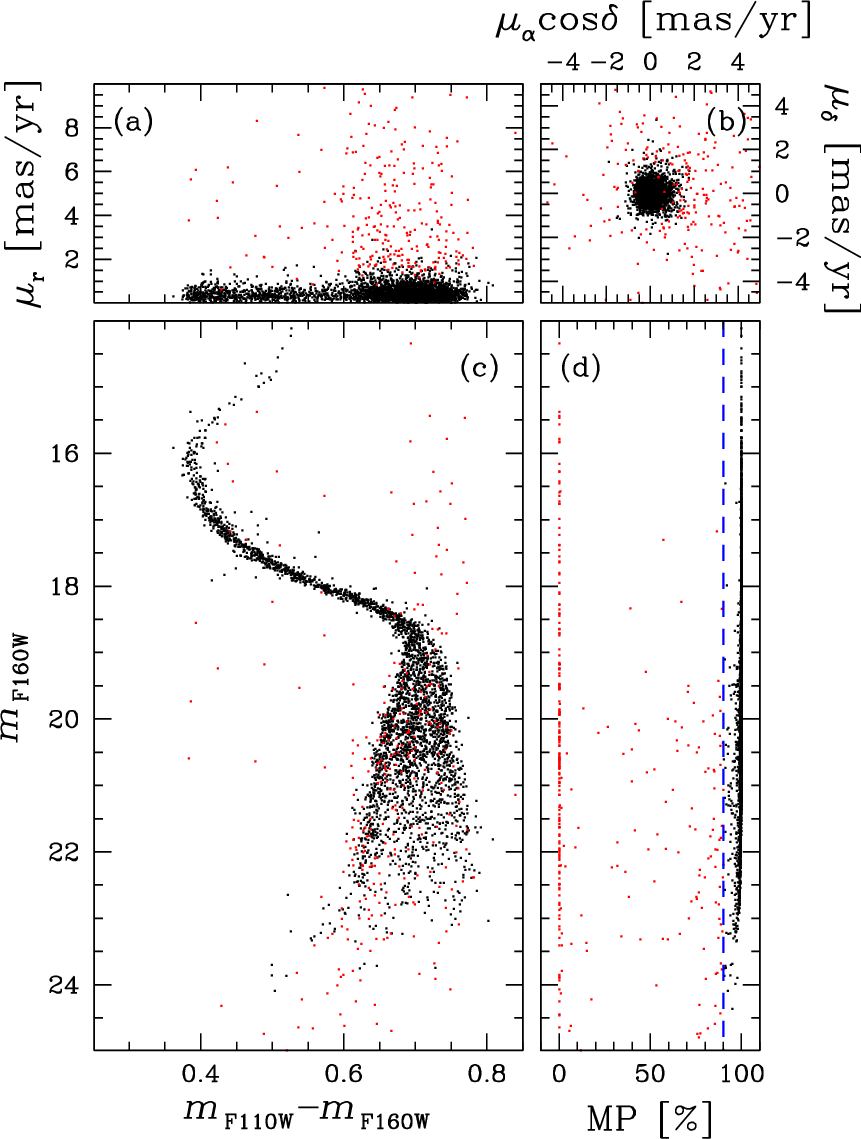}
 \caption{Cluster membership selection. (a) relative PMs as a function of the $m_{\rm F110W}-m_{\rm F160W}$ colour. (b) Vector point diagram. (c) $m_{\rm F160W}$ versus $m_{\rm F110W}-m_{\rm F160W}$ CMD. (d) Membership probability (MP) as a function of the $m_{\rm F160W}$ magnitude. The red line at MP=90\% denotes the MP threshold used in this study. In all panels, cluster members are represented in black, while field stars are shown in red.} 
 \label{Figure2} 
\end{figure}

We computed the membership probability (MP) of each star, by following a method based on PMs described by \citet{1998A&AS..133..387B} (see also \citealt{2009A&A...493..959B,2018MNRAS.481.3382N,2021MNRAS.505.3549S}). Figure\,\ref{Figure2} illustrates the selection of cluster members based on MP. We defined as cluster members all the sources with MP>90\% (as represented in panel\,(d) of Fig.\,\ref{Figure2}). Relative PMs as a function of the $m_{\rm F110W}-m_{\rm F160W}$ colour are shown in panel\,(a) while the vector-point diagram (VPD) is shown in panel\,(b). panel\,(c) shows the $m_{\rm F160W}$ versus $m_{\rm F110W}-m_{\rm F160W}$ CMD.

\begin{figure*}
 \centerline{\includegraphics[width=\textwidth]{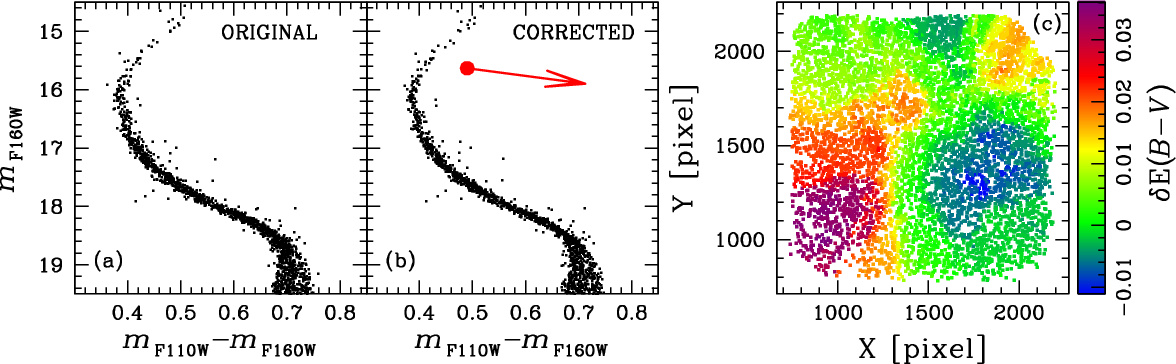}}
 \caption{Comparison between the original $m_{\rm F160W}$ versus $m_{\rm F110W}-m_{\rm F160W}$ CMD (a) and the same CMD corrected for differential reddening (b). The red arrow in (b) shows the reddening direction. (c) Differential reddening map, color-coded using the color-mapping scheme shown on the right.}
 \label{Figure3}
\end{figure*}

Next, we corrected our photometry for the effects of differential reddening on zero-point variations with location in the field following the procedure described by \citet{2012A&A...540A..16M} (see also \citealt{2007AJ....133.1658S,2017ApJ...842....7B}). In brief, we derived the fiducial MS line and measured the residual between a sample of bright MS stars and the fiducial along the reddening directions. For each star, we considered the median of the residual values from the 50 neighbouring bright MS stars as the best estimate of the differential reddening. Panels (a) and (b) of Fig.\,\ref{Figure3} provide a comparison of the $m_{\rm F160W}$ versus $m_{\rm F110W}-m_{\rm F160W}$ CMD in the upper part of the MS, where the effects of differential reddening are more pronounced, thanks to the negligible random errors. The figure illustrates the CMD before and after applying the differential-reddening correction. The values in magnitudes of the applied differential reddening ($\delta$E($B-V$)) of each star as a function of the (X, Y) stellar position can be used to construct the two-dimensional map of the differential reddening across the field of view (FoV). This map is shown in panel (c) of Fig.\,\ref{Figure3}, color-coded using the color-mapping scheme shown on the right of the figure.

\begin{figure*}
 \centerline{\includegraphics[width=\textwidth]{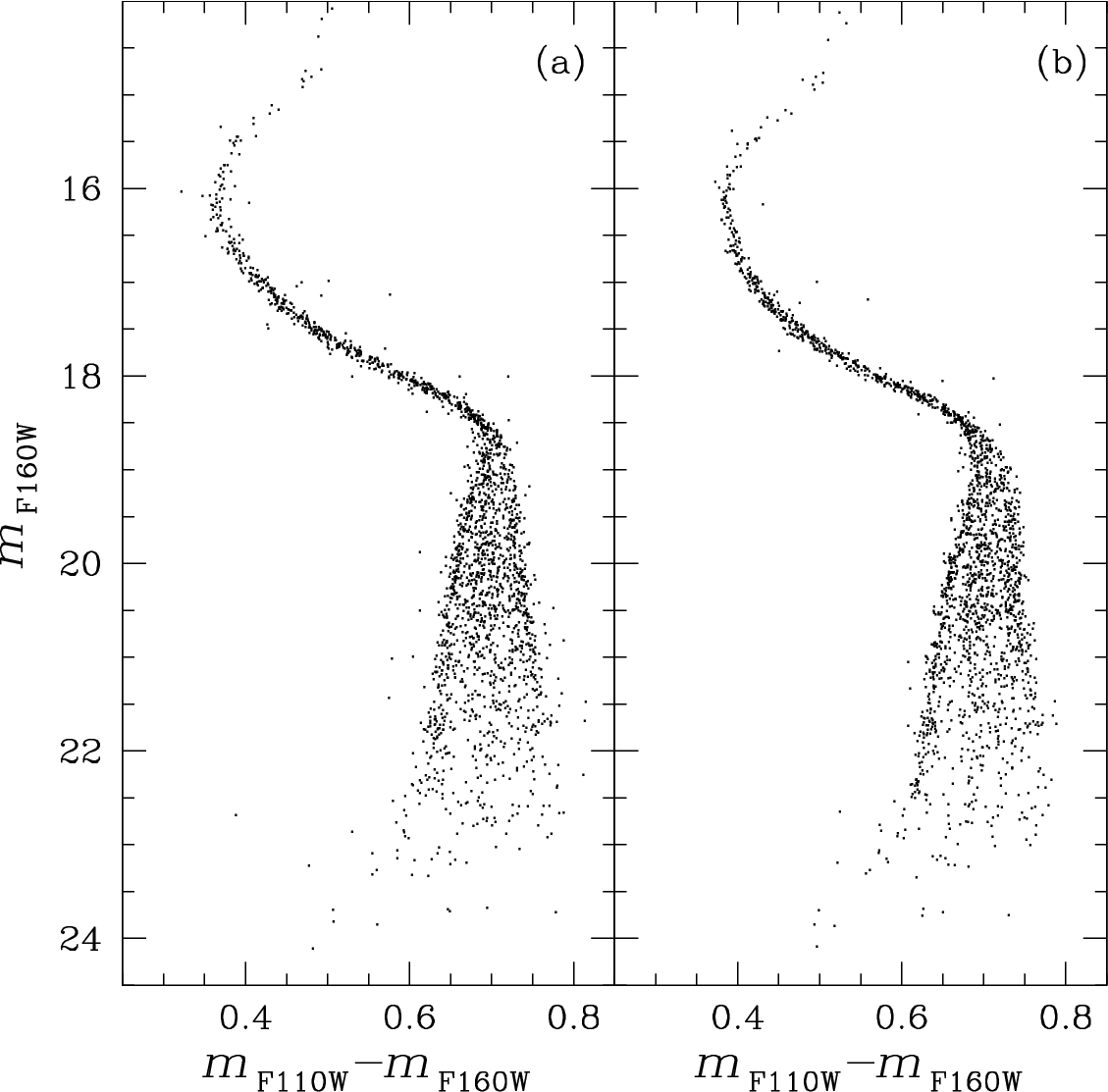}}
 \caption{Comparison between two CMDs made using the catalogue presented in \citetalias{2019MNRAS.484.4046M} (panel a) and our new reduction (panel b).} 
 \label{Figure4}
\end{figure*}

We cross-matched the stars that survived our photometric and membership selections with the catalogue presented in \citetalias{2019MNRAS.484.4046M}. Figure\,\ref{Figure4} shows the comparison between the two data sets, showing a significant improvement in the new catalogue. Notably, not only has the number of sources increased due to the expanded observational area, but stars in common with \citetalias{2019MNRAS.484.4046M} now benefit from twice the exposure time and a considerably larger time baseline. This has allowed us to derive improved PMs and, consequently, more accurate MPs for these stars. 

As part of this publication, we are making publicly available the astro-photometric catalogue and atlases we have constructed here. These resources are provided in a format identical to the catalogue and atlases made available by \citet{2021MNRAS.505.3549S}. For a comprehensive description of these resources, we refer to \citet{2021MNRAS.505.3549S}. When available, we also provide for sources in our catalogue the corresponding \textit{Gaia}\,DR3 identification numbers. The supplementary electronic material through this journal will also be available at our website\footnote{\href{https://web.oapd.inaf.it/bedin/files/PAPERs_eMATERIALs/NGC6752_V/}{https://web.oapd.inaf.it/bedin/files/PAPERs$\_$eMATERIALs/\\NGC6752$\_$V/}}.

\section{The Infrared CMD of NGC\,6752}\label{Section3}

\begin{figure*}
\centerline{\includegraphics[width=\textwidth]{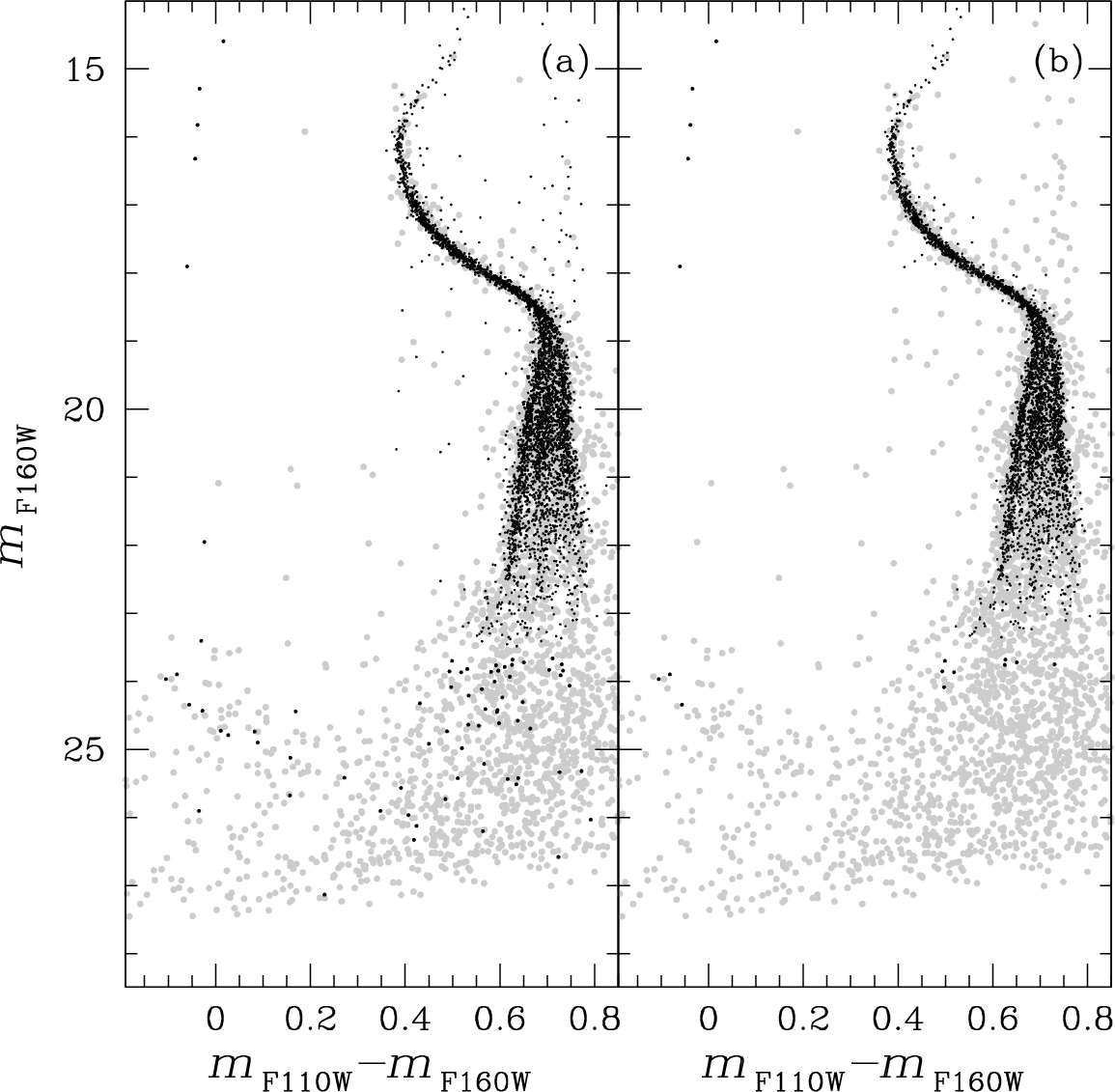}}
\caption{$m_{\rm F160W}$ versus $m_{\rm F110W}-m_{\rm F160W}$ CMD, illustrating stars that have passed the photometric selections (black points in panel\,(a)) and those that have passed both the photometric and MP selections (black points in panel\,(b)). Grey points in both panels represent the remaining stars. For enhanced visibility, we have emphasized the black points in the figure where $m_{\rm F110W}-m_{\rm F160W}$<0.3 or $m_{\rm F160W}$>23.5.} 
\label{Figure8.1} 
\end{figure*}

Figure\,\ref{Figure8.1} shows the $m_{\rm F160W}$ versus $m_{\rm F110W}-m_{\rm F160W}$ CMD of our catalogue. In panel\,(a), black points represent stars that have passed the photometric selections, whereas in panel\,(b), black points represent stars that have passed both the photometric and MP selections. Grey points in both panels represent the remaining stars.

Upon visual examination of this figure, it becomes evident that the MS is narrow and well-defined in the magnitude range between the MS turn-off ($m_{\rm F160W}\sim16$) and the MS knee ($m_{\rm F160W}\sim18.5$). However, as we move from the knee towards the lower end of the MS ($m_{\rm F160W}\sim24$), the MS becomes wider, and at this point, it becomes possible to distinguish the presence of three MS sequences of low-mass stars. The split of the MS is a direct outcome of the distinct oxygen abundance in the three populations \citepalias{2019MNRAS.484.4046M}. As discussed in Section\,\ref{Section1}, the $m_{\rm F110W}-m_{\rm F160W}$ colour is sensitive to the absorption bands of H$_2$O molecules, making it effective in distinguishing stars with different oxygen content. Following the nomenclature defined in \citetalias{2019MNRAS.484.4046M}), we label stars along the bluest sequence as MS-A, those along the reddest sequence (displaying the most significant oxygen depletion) as MS-C, and stars along the intermediate sequence as MS-B (showing an intermediate chemical composition). MS-A corresponds with the 1P stars, whereas MS-B and MS-C correspond to the 2Ps stars.

In what follows, we present an overview of our procedures for assessing the completeness of our sample, as well as our methodology for evaluating the isochrones of each population.  

\subsection{Artificial stars}\label{Section3.1}

\begin{figure*}
 \centerline{\includegraphics[width=\textwidth]{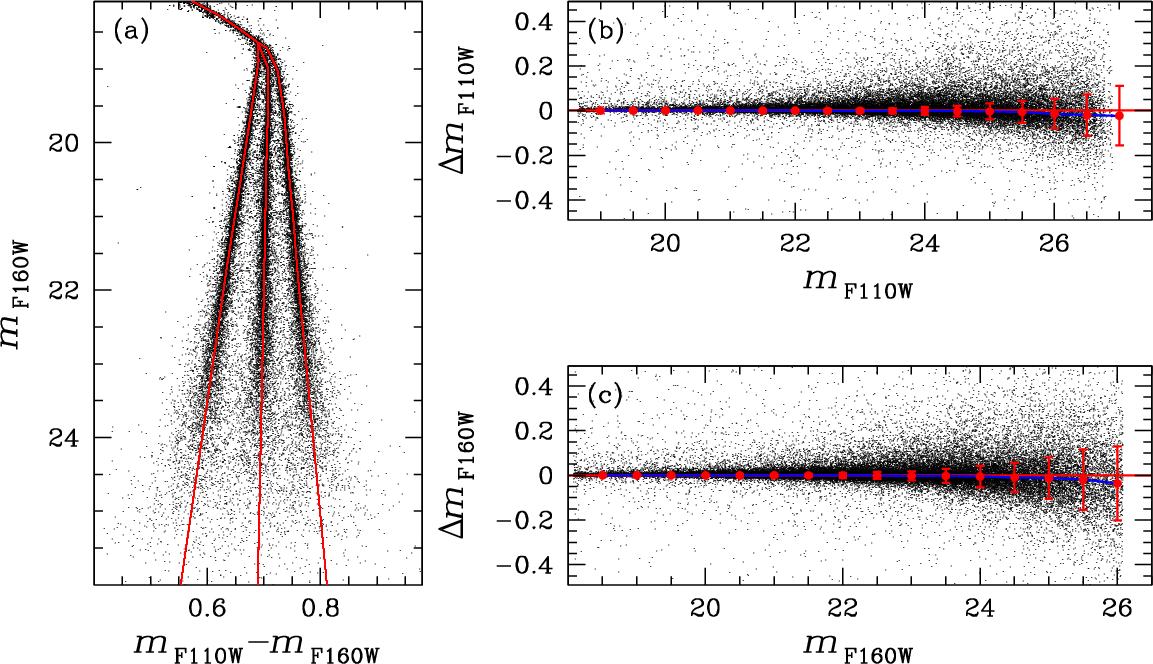}}
 \caption{(a) Simulated CMD derived from ASs. Red dots are the input ASs, while the measured ASs are coloured black. (b)-(c) Difference between inserted and recovered ASs magnitudes for F110W and F160W filter, respectively. We considered only stars found within a 0.5-pixel tolerance radius of the input coordinates. The red horizontal lines indicate the position of a null difference, while the blue lines are the splines through the median values obtained in each magnitude interval, represented with red points, with the corresponding standard deviation as red bars.} 
 \label{Figure5} 
\end{figure*}

Artificial stars (ASs) were used to estimate photometric errors and to derive the completeness level of our sample by following the procedure by \citet{2008ApJ...678.1279B,2009ApJ...697..965B}. We generated a catalogue including the X and Y positions and the F110W and F160W magnitudes of 90,000 stars (30,000 for each sequence) randomly distributed around the FoV. The ASs have instrumental magnitudes ranging from $-$6.5 to 1.5 (which correspond to $\sim18$ to $\sim26$ in calibrated magnitude) in the F160W band, while the corresponding F110W magnitudes are derived from the fiducial lines of the three sequences. These fiducial lines were established by manually setting a series of fiducial points along the three sequences in the $m_{\rm F160W}$ versus $m_{\rm F110W}-m_{\rm F160W}$ CMD and connecting them through straight lines (see panel\,(a) of Fig.\,\ref{Figure5}). The ASs were generated and reduced using the same software used for real stars, \texttt{KS2}. 

To track potential systematic errors, panels\,(b) and (c) of Fig.\,\ref{Figure5} show the difference between input and output (I/O) magnitudes versus input magnitude for ASs found within 0.5\,pixels of the input coordinates, in the F110W and F160W filter, respectively. For each filter, we divided the magnitude range covered by our AS photometry into bins of half magnitude and calculated the 2.5\,$\sigma$-clipped median of the differences per magnitude bin. The obtained values are plotted in Fig.\,\ref{Figure5} as red points with the corresponding standard deviation as error bars. Upon visual examination of this figure, it becomes evident that the mean differences are negligible. Consequently, based on the small magnitude of these differences, we have opted not to apply any input-output photometric correction. 

As pointed out in \citet{2008ApJ...678.1279B}, our ability to find a given faint star depends strongly on its environment. Extremely bright stars create a mottled background, making it hard to find faint sources in their vicinity.  For this reason, \citet{2008ApJ...678.1279B} introduced a new parameter, rmsSKY, which can be used to determine which locations in the field have a favourable background.  We then can define two types of completeness: an "overall completeness", $c$, defined as the fraction of stars recovered, and a "local completeness" defined as $c_g = c/f_g$, where $f_g$ is the fraction of the image where a source is searchable. To estimate $f_g$, we followed the procedure outlined in \citet{2008ApJ...678.1279B}. We plotted the rmsSKY parameter as a function of magnitude and defined by hand a line that separates well-measured stars from poorly measured ones. $f_g$ is defined as the fraction of stars that survived the rmsSKY selection criteria as a function of magnitude.

\begin{figure*} 
 \centerline{\includegraphics[width=\textwidth]{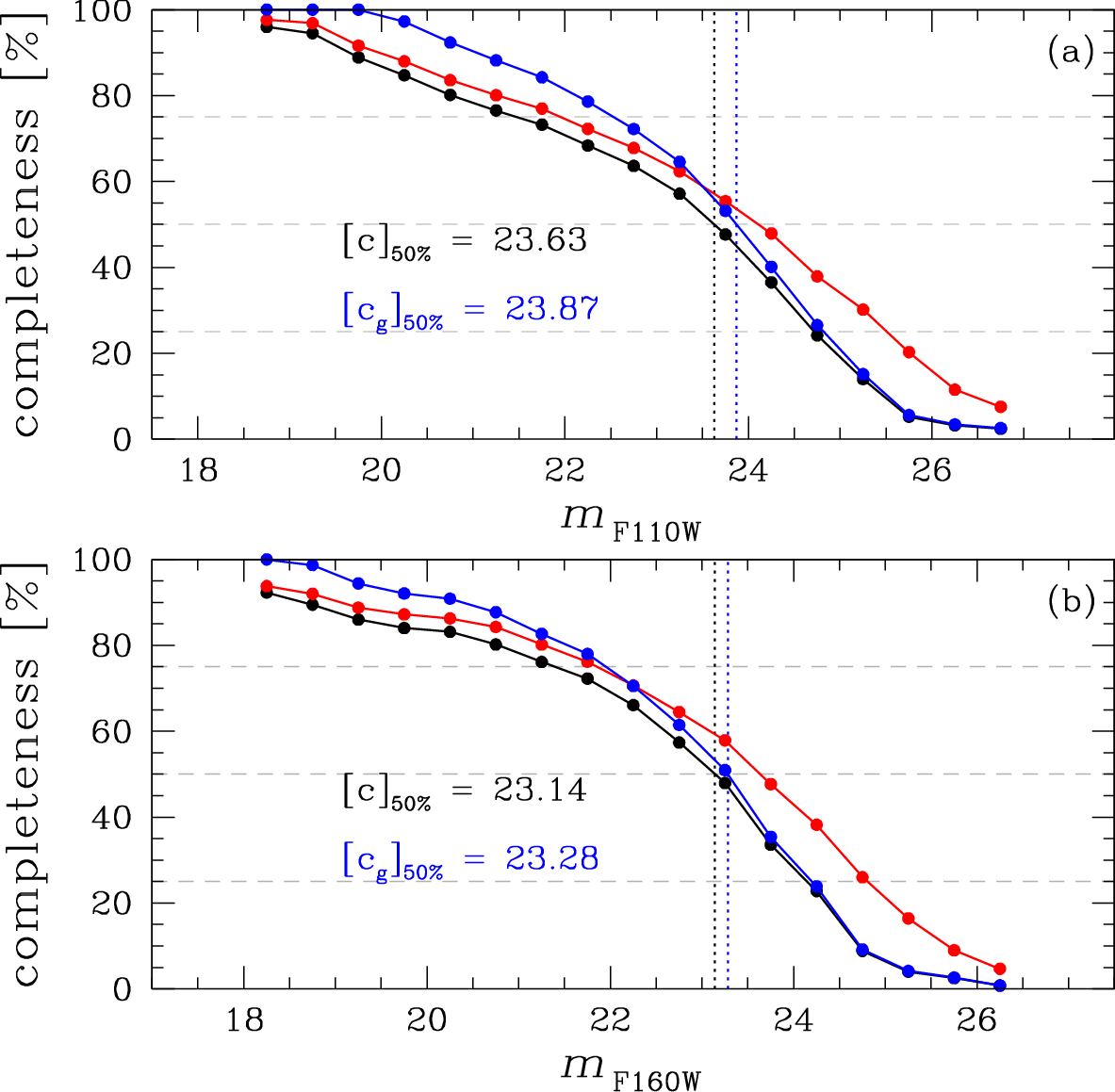}}
 \caption{Completeness based on AS tests for the F110W (a) and F160W (b) filter. Black points show the conventional completeness $c$, while the blue points are the low-rmsSKY completeness $c_g$. We also report, for each filter, the magnitude where $c$ and $c_g$ are equal to 50\% ($[c]_{50\%}$ and $[c_g]_{50\%}$). For comparison, we also show the conventional completeness obtained without considering the MP selection, represented with red points.}
 \label{Figure6}
\end{figure*}

Figure\,\ref{Figure6} shows the "overall" and "local" completeness level of our field, for the F110W (panel\,(a)) and F160W filter (panel\,(b)). For each filter, we divided the magnitude range covered by our AS photometry into bins of half magnitude and evaluated the percentage of AS sources recovered. An AS is considered recovered if the difference between the input and output position and magnitude is less than 1.0 \,pixel and 0.75\,mag. respectively, and if it passes the criteria of selection adopted for real stars (see Fig.\,\ref{Figure1}). To assess the impact of the MP selection (as illustrated in Fig.\,\ref{Figure2}) on completeness, we replicated the procedure for evaluating PMs and MPs, as applied to real stars, to the ASs. Specifically, we utilized the position of the ASs in each epoch, extracted through our ASs test, to assess the PMs of each AS by applying the same methodology used for the real stars. Subsequently, we computed the MP for the ASs using the obtained PMs, following the procedure described above for real stars. We applied identical MP selection criteria to the ASs as those used for the real stars. 

The "overall" and "local" completeness estimates are represented with black and blue dots respectively in both panels. For comparison, we also show the "overall" completeness obtained without considering the MP selection, represented by red dots in both panels. As expected, completeness is higher in the absence of MP selection, particularly at lower magnitudes where obtaining reliable PM estimates can be challenging. The incorporation of PMs in the completeness evaluation represents a significant improvement compared to \citetalias{2019MNRAS.484.4046M}. We want to highlight that, in the upcoming analysis, we will rely solely on the "overall" completeness obtained with all the selections. 

\begin{figure*}
 \centerline{\includegraphics[width=\textwidth]{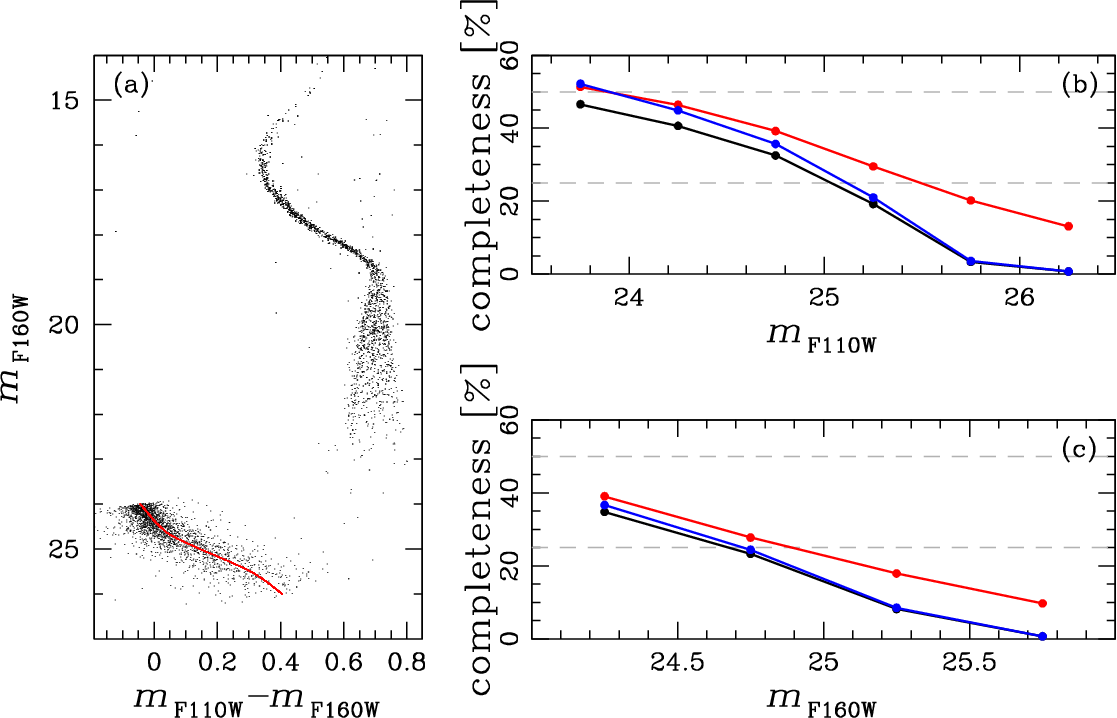}}
 \caption{Same as panel\,(a) of Fig.\,\ref{Figure5} and panels\,(a) and (b) of Fig.\,\ref{Figure6} but for the WD CS.} 
 \label{Figure10} 
\end{figure*}

We followed a similar procedure to assess the completeness of stars along the WD CS. Specifically, we generated a sample comprising 15,000 stars with instrumental magnitudes spanning from $-$0.5 to 1.5 (which correspond to $\sim24$ to $\sim26$ in calibrated magnitude) in the F160W band. The corresponding F110W magnitudes were derived from the fiducial line of the WD CS in the $m_{\rm F160W}$ versus $m_{\rm F110W}-m_{\rm F160W}$ CMD (refer to panel\,(a) of Fig.\,\ref{Figure10}). In Fig.\,\ref{Figure10}, panels\,(b) and (c) illustrate the completeness levels for stars along the WD CS. The "overall" completeness level, with MP selection, is represented by black dots, while completeness without MP selection is indicated by red dots. Additionally, the "local" completeness levels are depicted as blue dots. It's worth noting that the completeness level remains below 50\% across the entire magnitude range under consideration. 

The completeness values for both the "overall" and "local" completeness of our sample in the two filters, as well as for both the MS and the WD CS, can be accessed online, on our website and as supplementary electronic material within this journal (alongside the catalogue and the atlases).

\subsection{Isochrone fits}\label{Section3.2}

Model isochrones were calculated and fit to the observed CMD of NGC\,6752, following the approach in \citet{roman_theory_paper}. This approach allows the abundances of individual elements to be adjusted to match observations as well as possible.  The effect of non-solar element abundances is thoroughly accounted for in stellar interiors and atmospheres, as well as the boundary conditions connecting them. The isochrones are based on new evolutionary models, calculated using the \texttt{MESA} code \citep{MESA,MESA_2,MESA_3,MESA_4,MESA_5}, and new model atmospheres, calculated using \texttt{PHOENIX 15} \citep{phoenix_origin,BT-Settl,phoenix_15,roman_note} and \texttt{BasicATLAS / ATLAS 9} \citep{ATLAS5,ATLAS9_1,ATLAS9_2,mikaela}.

The parameters of our isochrones are summarised in Table\,\ref{tab:iso_params}. Among the spectroscopic abundances in \citet{2013ApJ...767..120M} and \citet{yong_2005,yong_2008} that were used as the initial guesses in our analysis, $\mathrm{[O/Fe]}$, $\mathrm{[Na/Fe]}$, $\mathrm{[Al/Fe]}$ needed to be adjusted to reconcile our model isochrones with the lower MS photometry. Additionally, non-solar $\mathrm{[C/Fe]}$ values needed to be adopted for MS-A and MS-B. The adjusted photometric abundances retain the key features of GC chemistry, specifically the $\mathrm{Na}$ -- $\mathrm{O}$ and $\mathrm{N}$ -- $\mathrm{C}$ anti-correlations; however, we find that the difference between 1P and 2P abundances that were required to reproduce the photometric scatter is smaller than the spectroscopic abundance spread of red giants by $0.2-0.3\,\mathrm{dex}$. Our photometric estimate of $\mathrm{[Al/Fe]}$ appears to consistently exceed the spectroscopic counterpart in all three populations by up to $0.4\,\mathrm{dex}$. Significant discrepancies between photometric and spectroscopic abundances are not unexpected, since both methods are affected by distinct systematic errors. The offset between spectroscopic and photometric estimates of $\mathrm{[O/Fe]}$ of order $0.15\,\mathrm{dex}$ shown in Table\,\ref{tab:iso_params} is consistent with previous work \citep{roman_theory_paper}. The large offset in $\mathrm{[Al/Fe]}$ is likely a combination of systematic errors in both photometric and spectroscopic estimates due to the poorer isochrone fit in the vicinity of the MS knee, where the photometric constraints on $\mathrm{[Al/Fe]}$ are strictest, and strong non-local thermodynamic equilibrium effects on the spectral features of aluminium \citep{NLTE_Al}. In general, small offsets between spectroscopic and photometric abundances indicate that the quoted values are particularly reliable, while larger offsets require a follow-up investigation of associated systematic errors. In particular, more robust photometric abundances may be derived by considering the colours in additional photometric bands.

All model atmospheres, evolutionary models, and isochrones employed in this study are available online\footnote{\href{http://romanger.com/models.html}{http://romanger.com/models.html}}.

\begin{table*}
\caption{Parameters of new NGC\,6752 isochrones calculated in this study. $\alpha_\mathrm{MLT}$ is the assumed convective mixing length in the interior in terms of pressure scale heights. $(m-M)_0$ and $E(B-V)$ are the distance modulus and optical reddening in $\mathrm{mag}$, respectively. $Y$ is the helium mass fraction. All abundances are given in $\mathrm{dex}$ with respect to the solar standard used in \citet{roman_omega_cen}. The values estimated from isochrone fitting are highlighted in bold; the rest of the parameters were adopted from the literature. In the ``Ref.'' columns, ``M19'' refers to \citetalias{2019MNRAS.484.4046M}; ``M13'' to the compilation of spectroscopic abundances in Table\,2 of \citet{2013ApJ...767..120M}, derived from \citet{yong_2005,yong_2008}; ``M13Y'' to the best-fit $Y$ values for the three populations from \citet{2013ApJ...767..120M}; ``C10'' to the recommended solar abundance of carbon from \citet{C10}, ``TO'' to the isochrone best fit around the MS turn-off point and the subgiant branch; and ``LMS'' to the isochrone best fit near the end of the MS below the MS knee. The ``Offset'' columns show the difference between the photometric best fit and the spectroscopic value from \citet{2013ApJ...767..120M}, where applicable. Age, $\alpha_\mathrm{MLT}$, $(m-M)_0$ and $E(B-V)$ were assumed to be identical for all three populations. The errors in ``TO'' parameters are taken directly from the covariance matrix of the fit. The errors in ``LMS'' parameters are taken as the weighted standard deviation of the best-fit value along the lower MS.}
\centering
\label{tab:iso_params}
\begin{tabular}{l|lcc|lcc|lcc}
    \hline
     & \multicolumn{3}{c|}{MS-A} & \multicolumn{3}{c|}{MS-B} & \multicolumn{3}{c}{MS-C} \\
     \hline
    Parameter & Value & Offset & Ref. & Value & Offset & Ref. & Value & Offset & Ref. \\
    \hline
    $\alpha_\mathrm{MLT}$ & $\pmb{1.57}$ $\pm0.05$ & $-$ & TO & $\pmb{1.57}$ $\pm0.05$ & $-$ & TO & $\pmb{1.57}$ $\pm0.05$ & $-$ & TO \\
    Age [$\mathrm{Gyr}$] & $\pmb{12.63}$ $\pm0.40$ & $-$ & TO & $\pmb{12.63}$ $\pm0.40$ & $-$ & TO & $\pmb{12.63}$ $\pm0.40$ & $-$ & TO \\
    $(m-M)_0$ & $\pmb{13.10}$ $\pm0.01$ & $-$ & TO & $\pmb{13.10}$ $\pm0.01$ & $-$ & TO & $\pmb{13.10}$ $\pm0.01$ & $-$ & TO \\
    $E(B-V)$ & $+0.07$ & $-$ & M19 & $+0.07$ & $-$ & M19 & $+0.07$ & $-$ & M19 \\
    $Y$ & $0.246$ & $-$ & M13Y & $0.254$ & $-$ & M13Y & $0.275$ & $-$ & M13Y \\
    $[\mathrm{Fe/H}]$ & $-1.65$ & $-$ & M13 & $-1.61$ & $-$ & M13 & $-1.61$ & $-$ & M13 \\
    $[\mathrm{C/Fe}]$ & $\pmb{+0.20}$ $\pm0.19$ & $-$ & LMS & $\pmb{-0.30}$ $\pm0.19$ & $-$ & LMS & $+0.00$  & $-$ & C10 \\
    $[\mathrm{N/Fe}]$ & $-0.11$  & $-$ & M13 & $+0.92$  & $-$ & M13 & $+1.35$  & $-$ & M13 \\
    $[\mathrm{O/Fe}]$ & $\pmb{+0.67}$ $\pm0.03$ & $+0.02$ & LMS & $\pmb{+0.50}$ $\pm0.04$ & $+0.07$ & LMS & $\pmb{+0.23}$ $\pm0.02$ & $+0.20$ & LMS \\
    $[\mathrm{Na/Fe}]$ & $-0.03$  & $-$ & M13 & $+0.26$  & $-$ & M13 & $\pmb{+0.31}$ $\pm0.05$ & $-0.30$ & LMS \\
    $[\mathrm{Mg/Fe}]$ & $+0.51$  & $-$ & M13 & $+0.49$  & $-$ & M13 & $+0.40$  & $-$ & M13 \\
    $[\mathrm{Al/Fe}]$ & $\pmb{+0.65}$ $\pm0.13$ & $+0.37$ & LMS & $\pmb{+1.10}$ $\pm0.04$ & $+0.40$ & LMS & $\pmb{+1.34}$ $\pm0.03$ & $+0.20$ & LMS \\
    $[\mathrm{Si/Fe}]$ & $+0.27$  & $-$ & M13 & $+0.33$  & $-$ & M13 & $+0.35$  & $-$ & M13 \\
    $[\mathrm{Ca/Fe}]$ & $+0.21$  & $-$ & M13 & $+0.24$  & $-$ & M13 & $+0.27$  & $-$ & M13 \\
    $[\mathrm{Sc/Fe}]$ & $-0.05$  & $-$ & M13 & $-0.04$  & $-$ & M13 & $-0.04$  & $-$ & M13 \\
    $[\mathrm{Ti/Fe}]$ & $+0.10$  & $-$ & M13 & $+0.14$  & $-$ & M13 & $+0.15$  & $-$ & M13 \\
    $[\mathrm{V/Fe}]$ & $-0.34$  & $-$ & M13 & $-0.29$  & $-$ & M13 & $-0.25$  & $-$ & M13 \\
    $[\mathrm{Mn/Fe}]$ & $-0.50$  & $-$ & M13 & $-0.44$  & $-$ & M13 & $-0.45$  & $-$ & M13 \\
    $[\mathrm{Co/Fe}]$ & $-0.03$  & $-$ & M13 & $-0.00$  & $-$ & M13 & $-0.06$  & $-$ & M13 \\
    $[\mathrm{Ni/Fe}]$ & $-0.06$  & $-$ & M13 & $-0.06$  & $-$ & M13 & $-0.03$  & $-$ & M13 \\
    $[\mathrm{Cu/Fe}]$ & $-0.66$  & $-$ & M13 & $-0.59$  & $-$ & M13 & $-0.60$  & $-$ & M13 \\
    $[\mathrm{Y/Fe}]$ & $-0.09$  & $-$ & M13 & $-0.01$  & $-$ & M13 & $+0.01$  & $-$ & M13 \\
    $[\mathrm{Zr/Fe}]$ & $+0.07$  & $-$ & M13 & $+0.20$  & $-$ & M13 & $+0.21$  & $-$ & M13 \\
    $[\mathrm{Ba/Fe}]$ & $-0.09$  & $-$ & M13 & $-0.12$  & $-$ & M13 & $+0.05$  & $-$ & M13 \\
    $[\mathrm{La/Fe}]$ & $+0.12$  & $-$ & M13 & $+0.10$  & $-$ & M13 & $+0.13$  & $-$ & M13 \\
    $[\mathrm{Ce/Fe}]$ & $+0.28$  & $-$ & M13 & $+0.25$  & $-$ & M13 & $+0.28$  & $-$ & M13 \\
    $[\mathrm{Nd/Fe}]$ & $+0.23$  & $-$ & M13 & $+0.22$  & $-$ & M13 & $+0.23$  & $-$ & M13 \\
    $[\mathrm{Eu/Fe}]$ & $+0.31$  & $-$ & M13 & $+0.30$  & $-$ & M13 & $+0.34$  & $-$ & M13 \\
\hline
\end{tabular}
\end{table*}

\begin{figure*}
\centerline{\includegraphics[width=\textwidth]{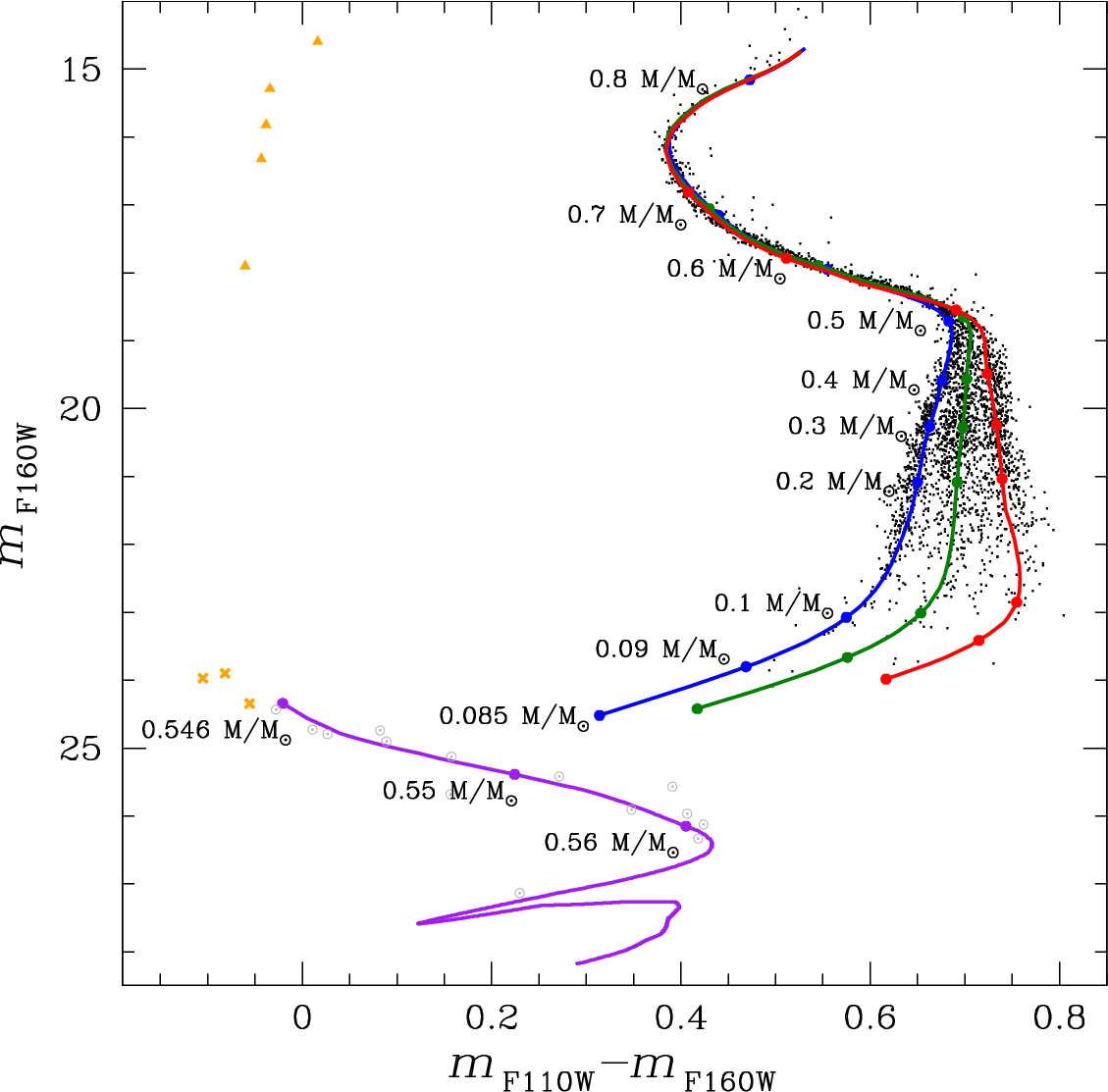}}
\caption{The three isochrones described in Section \ref{Section3.2} corresponding to MS-A (in blue), MS-B (in green), and MS-C (in red) super-imposed on the $m_{\rm F160W}$ versus $m_{\rm F110W}-m_{\rm F160W}$ CMD of the stars that have passed both the photometric and MP selections. The corresponding isochrone for the WD is shown in purple in the lower-left part of the plot, with the three WD candidates shown as orange crosses. Other stars that only passed the photometric selections and are situated along the WD isochrone are enclosed by a grey circle. We also show, as orange triangles, stars belonging to the horizontal branch (HB). For each isochrone, we've marked specific mass values, with their labels displayed on the plot.} 
\label{Figure8.2} 
\end{figure*}

The final fits of three isochrones corresponding to MS-A, MS-B, and MS-C are represented in Fig.\,\ref{Figure8.2}, in blue, green, and red, respectively. In addition to the isochrones, we've marked specific mass values
with their corresponding labels on the plot.

\begin{figure}
\centerline{\includegraphics[width=\columnwidth]{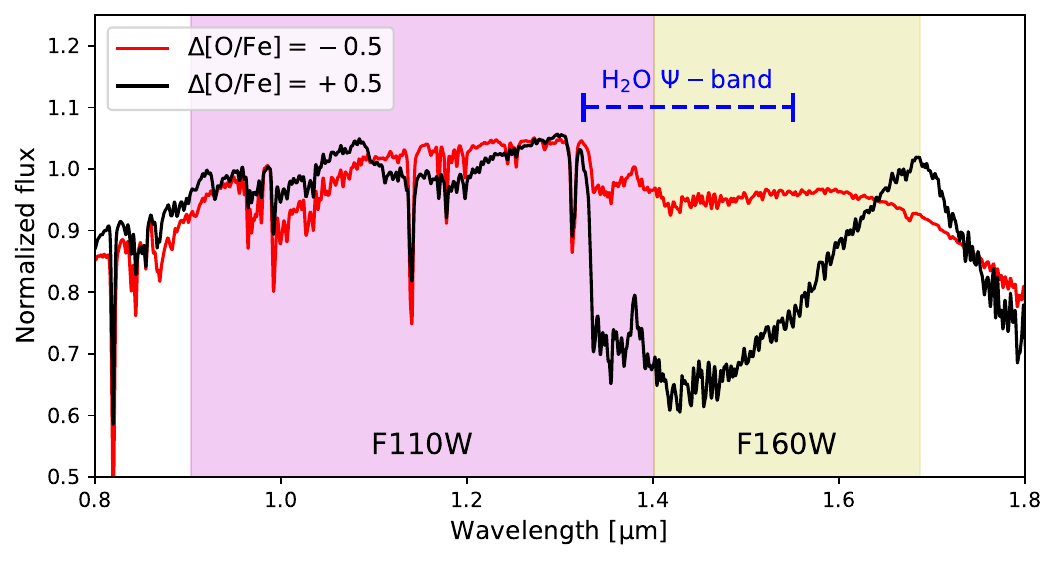}}
\caption{Synthetic spectra for two stars from MS-B and with stellar mass $0.095$\,M$_{\odot}$, characterised by different content of $[\mathrm{O/Fe}]$. The wavelength windows of the WFC3/NIR F110W and F160W filter bands are highlighted in magenta and yellow colours, respectively. We also denote the H$_2$O absorption band.} 
\label{spectra} 
\end{figure}

To illustrate the impact of oxygen variations on the \textit{HST} NIR bands, Fig.\,\ref{spectra} shows two synthetic spectra representing stars from MS-B and stellar mass $0.095$\,M$_{\odot}$. These two spectra are characterised by different content of $[\mathrm{O/Fe}]$, specifically $\Delta[\mathrm{O/Fe}]=-0.5$ (in red) and $\Delta[\mathrm{O/Fe}]=0.5$ (in black), with respect to the $[\mathrm{O/Fe}]$ value reported in Table \ref{tab:iso_params}. The figure focuses on the wavelength range covered by the WFC3/NIR F110W and F160W filter bands, highlighted by magenta and yellow colours, respectively. The H$_2$O absorption band is also indicated. The H$_2$O molecules have a strong effect on the two synthetic spectra, resulting in a notably lower flux 
in the F160W filter.

\subsubsection{The WD CS of NGC\,6752 in the infrared}\label{Section3.2.1}

A representative 12.5\,Gyr WD isochrone is shown in the lower-left part of Fig.\,\ref{Figure8.2}. It has been computed using the WD models by \citet{bastiwd} calculated for metal-poor progenitors employing the \citet{elopa} electron conduction opacities \citep[see][for more details]{bastiwd}, the WD initial-final mass relation by \citet{ifmr}, and progenitor lifetimes taken from the $\alpha$-enhanced models by \citet{bastialpha}.

Notably, in the upper part of the WD isochrone, we observe three sources that survived both the photometric and MP selections (shown as orange crosses) and lie very close to the isochrone. Other stars that passed only the photometric selections are situated along the WD isochrone (marked by a grey circle around them), making them potential WD candidates. However, due to the limited depth of our data, we cannot ensure a sufficient level of PM accuracy to distinguish reliably between cluster members and field stars in this faint region of the CMD. Consequently, more data will be necessary to confirm the presence of an extended WD sequence.

We employed the WD isochrone shown in the plot to determine the effective temperature ($T_{\rm eff}$) and gravity ($g$) values for the three surviving WDs, which passed both the photometric and MP selection criteria. The resulting values are as follows, ordered from the brightest to the faintest source: $T_{\rm eff}$ = $(2.510\pm0.223)\times10^{4}$\,K, $(2.379\pm0.212)\times10^{4}$\,K and $(1.782\pm0.159)\times10^{4}$\,K, and log($g$) = 7.824, 7.829 and 7.857, in cgs units.

\section{Luminosity Function of sub-populations}\label{Section4}

In this Section, we present the LFs for the three MSs within the magnitude range of $19<m_{\rm F160W}<23.5$. Two different methods were used to assess the LFs, which are described in the following.

\subsection{Luminosity Function using regions in the verticalised CMD}\label{Section4.1}

\begin{figure*}
 \centerline{\includegraphics[width=\textwidth]{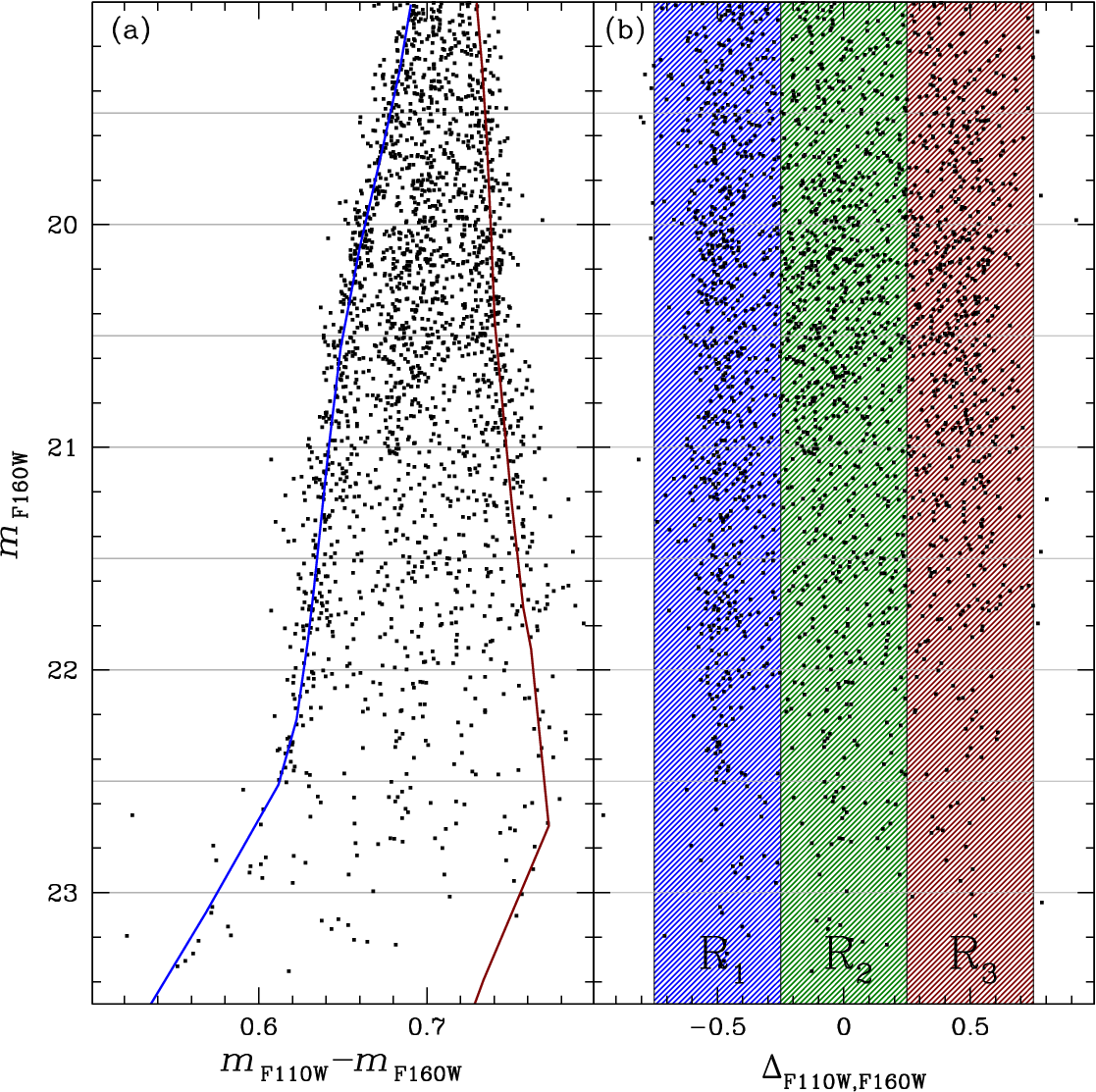}}
 \caption{Procedure for estimating the LFs of MS-A, MS-B, and MS-C. (a) zoomed view of the $m_{\rm F160W}$ versus $m_{\rm F110W}-m_{\rm F160W}$ CMD, focusing on the region where the three MSs are distinctly identifiable. The bluest and reddest sequences are associated with MS-A and MS-C stars, respectively, while the middle sequence corresponds to MS-B stars. The blue and red lines represent the fiducials of MS-A and MS-C, respectively, and are utilised to construct the verticalized CMD shown in (b). (b) in the verticalized CMD, three distinct regions, denoted as R$_1$, R$_2$, and R$_3$, are defined, each corresponding to one of the three MS populations (MS-A, MS-B, and MS-C), distinguished by blue, green, and red colours, respectively.} 
 \label{Figure7} 
\end{figure*}

The initial procedure follows a method similar to the one outlined in \citet{2012A&A...537A..77M}. In panel\,(a) of Fig.\,\ref{Figure7}, we present a zoomed region of the $m_{\rm F160W}$ versus $m_{\rm F110W}-m_{\rm F160W}$ CMD shown in Fig.\,\ref{Figure8.2}, where the triple MS sequence is most evident. 

The fiducial lines, constructed by connecting manually drawn fiducial points on the CMD with straight lines, appear as blue and red lines for MS-A and MS-C, respectively, on the CMD. These fiducial lines were used to create the diagram shown in panel\,(b) of Fig.\,\ref{Figure7}, with the abscissa calculated using equation\,(1) from \citetalias{2019MNRAS.484.4046M}.
In this diagram, we established three distinct regions, denoted as R$_1$, R$_2$, and R$_3$, each corresponding to the three MS populations, MS-A, MS-B, and MS-C. These defined regions are visually represented by blue, green, and red colours, respectively, in panel\,(b) of Fig.\,\ref{Figure7}.

It is important to note that not all the single MS-A(B,C) stars are confined solely within R$_{1(2,3)}$, as a fraction of them may migrate to nearby regions due to a combination of measurement errors and overlapping distributions of the chemical abundances within each MS. We will evaluate this latter effect in the following section. In this section, we're focusing solely on the impacts stemming from measurement errors. To account for this contamination effect and estimate the actual number of stars in each population we utilised the following procedure:

The total numbers $N_{i=1,2,3}$ (corrected for completeness) of stars within each region R$_{i=1,2,3}$ can be expressed as the sum of three terms:
\begin{equation}
    N_i = N_{\rm A} f_i^{\rm A} + N_{\rm B} f_i^{\rm B} + N_{\rm C} f_i^{\rm C}.
    \label{equation}
\end{equation}
where $N_{\rm A}$, $N_{\rm B}$, and $N_{\rm C}$ represent the total number of stars in the three populations, and $f_i^{\rm A}$, $f_i^{\rm B}$, and $f_i^{\rm C}$ are the fractions of stars from the three populations that fall within the $i$-th region\footnote{In the analysed region of the $m_{\rm F160W}$ versus $m_{\rm F110W}-m_{\rm F160W}$ CMD, the MSs exhibit a nearly vertical orientation, causing binary stars to be closely mixed with single stars. As a result, in this analysis, we did not consider the presence of binaries. However, we emphasise, that the observed binary fraction in this cluster is rather low anyway \citep[$\sim$0.7\%, see][]{2010ApJ...709.1183M}.}.

To estimate the quantities $f_i^{\rm A}$, $f_i^{\rm B}$, and $f_i^{\rm C}$, we employed ASs (see Section\,\ref{Section3.1}). For each real star in our catalogue, we identified the five closest ASs in terms of position and magnitude and defined a sample of ASs. We then verticalized the obtained ASs sample using the same procedure and fiducial lines shown in Fig.\,\ref{Figure7}. The values of $f_i^{\rm A}$, $f_i^{\rm B}$, and $f_i^{\rm C}$ were evaluated as the fraction of ASs associated with MS-A, MS-B, and MS-C stars that fall in each region R$_1$, R$_2$, and R$_3$.

\begin{figure*}
 \centerline{\includegraphics[width=\textwidth]{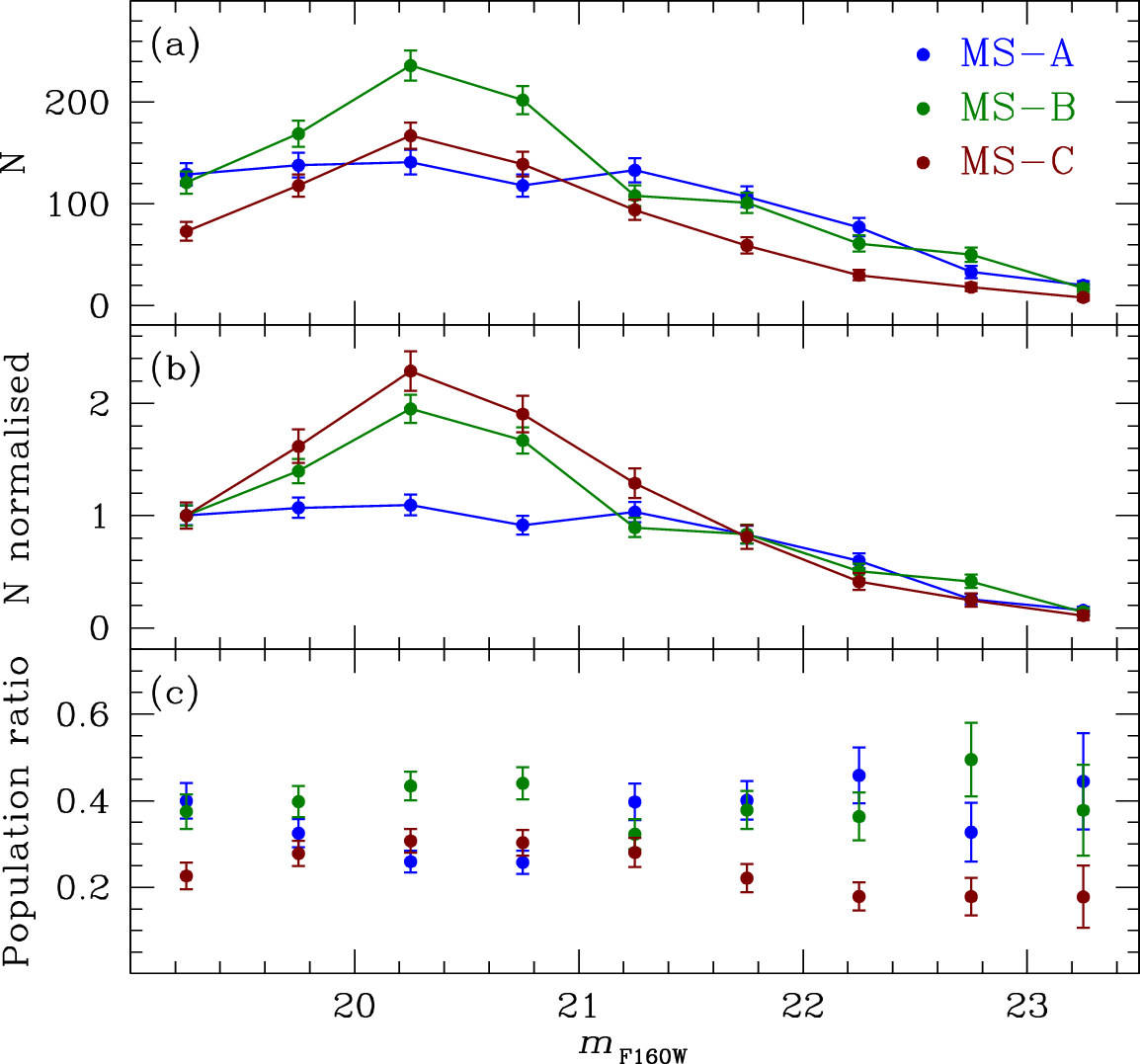}}
 \caption{LFs estimated using regions in the verticalised CMD (see Fig.\,\ref{Figure7}) (a) LFs for the three MSs. (b) LFs for the three MSs, normalised by the value of the first bin for each respective MS. (c) Number ratios for the three stellar populations with respect to the total, as a function of the $m_{\rm F160W}$ magnitude.} 
 \label{Figure9} 
\end{figure*}

We solved Eq.\,\ref{equation} for the nine magnitude bins defined by the grey horizontal lines in Fig.\,\ref{Figure7}, obtaining estimates of $N_{\rm A}$, $N_{\rm B}$, and $N_{\rm C}$ for each bin. The corresponding values are reported in Table\,\ref{Table4}, while the three LFs for the three individual MSs are displayed in panel\,(a) of Fig.\,\ref{Figure9}, with the associated errors represented as Poisson errors. Within the magnitude interval $19 < m_{\rm F160W} < 21.25$, MS-B exhibits the highest number of stars, while populations MS-A and MS-C have a smaller and similar number of stars. Moving to the magnitude interval $21.25 < m_{\rm F160W} < 23.5$, MS-B and MS-A are the most populated, with nearly the same number of stars, while MS-C shows a lower number of stars. An examination of this figure seems to reveal that MS-B and MS-C LFs follow a comparable trend with magnitude. They exhibit an increase up to approximately $m_{\rm F160W}\sim20.25$, followed by a gradual decrease for faint magnitudes. In contrast, the LF for MS-A remains relatively constant in the brighter region and only starts to decrease at magnitudes fainter than $m_{\rm F160W}\sim21.5$, with a behaviour similar to those of the LFs of MS-B and MS-C.  

These distinctions become even more pronounced in panel\,(b) of Fig.\,\ref{Figure9}, where the LFs for the three individual MSs are normalised by the value of the first bin for each respective MS. Notably, within the magnitude interval of $19 < m_{\rm F160W} < 21.25$, the three MSs manifest distinct trends, with MS-B and MS-C displaying a pronounced increase with respect to MS-A. In the magnitude range of $21.25 < m_{\rm F160W} < 23.5$, the three MSs share a similar trend.

Finally, panel\,(c) of Fig.\,\ref{Figure9} presents the ratio of $N_{\rm A}$, $N_{\rm B}$, and $N_{\rm C}$ to the total number of stars, $N$. The values are reported in Table\,\ref{Table3}.

\begin{table*}
\caption{Number of stars for each MS ($N_{\rm A}$, $N_{\rm B}$ and $N_{\rm C}$) in nine magnitude bins and for the two methods.}
\centering
 \label{Table4}
\begin{tabular}{c @{\quad \vline \quad} c c c @{\quad \vline \quad} c c c}
\hline
 $\Delta m_{\rm F160W}$ & & Regions & & & Gaussian &\\
\hline
  & $N_{\rm A}$ & $N_{\rm B}$ & $N_{\rm C}$ & $N_{\rm A}$ & $N_{\rm B}$ & $N_{\rm C}$\\
 \hline
19.0-19.5 & 129$\pm$11 & 121$\pm$11 &  73$\pm$9  & 124$\pm$11 &  68$\pm$8  & 126$\pm$11\\
19.5-20.0 & 138$\pm$12 & 169$\pm$13 & 118$\pm$11 & 112$\pm$11 & 180$\pm$13 & 128$\pm$11\\
20.0-20.5 & 141$\pm$12 & 236$\pm$15 & 167$\pm$13 & 100$\pm$10 & 288$\pm$17 & 156$\pm$12\\
20.5-21.0 & 118$\pm$11 & 202$\pm$14 & 139$\pm$12 &  97$\pm$10 & 215$\pm$15 & 146$\pm$12\\
21.0-21.5 & 133$\pm$12 & 108$\pm$10 &  94$\pm$10 & 100$\pm$10 & 177$\pm$13 &  59$\pm$8\\
21.5-22.0 & 107$\pm$10 & 101$\pm$10 &  59$\pm$8  &  63$\pm$8  & 189$\pm$14 &  17$\pm$4\\
22.0-22.5 &  77$\pm$9  &  61$\pm$8  &  30$\pm$5  &  58$\pm$8  &  81$\pm$9  &  25$\pm$5\\
22.5-23.0 &  33$\pm$6  &  50$\pm$7  &  18$\pm$4  &  31$\pm$6  &  38$\pm$6  &  31$\pm$6\\
23.0-23.5 &  20$\pm$4  &  17$\pm$4  &   8$\pm$3  &  19$\pm$4  &  11$\pm$3  &  15$\pm$4\\
\hline
 \end{tabular}
\end{table*}
\begin{table*}
\caption{Fractions of MS-A, MS-B, and MS-C stars relative to the total number of MS stars in nine $m_{\rm F160W}$ magnitude bins, obtained through the two different methods. Additionally, we provide the dispersions of the three best-fitting Gaussian functions ($\sigma_{\rm MS-A, MS-B, MS-C}$). We also provide the weighted mean values of each fraction, with the uncertainties estimated as the ratio between the root mean square (rms) of the nine population-ratio measurements divided by the square root of 8.}
\centering
 \label{Table3}
\begin{tabular}{c @{\quad \vline \quad} c c c @{\quad \vline \quad} c c c @{\quad \vline \quad} c c c}
 \hline
 $\Delta m_{\rm F160W}$ & & Regions & & & Gaussian & & & & \\
 \hline
  & $N_{\rm A}/N$ & $N_{\rm B}/N$ & $N_{\rm C}/N$ & $N_{\rm A}/N$ & $N_{\rm B}/N$ & $N_{\rm C}/N$ & $\sigma_{\rm MS-A}$ & $\sigma_{\rm MS-B}$ & $\sigma_{\rm MS-C}$\\
 \hline
19.0-19.5 & 0.40$\pm$0.04 & 0.37$\pm$0.04 & 0.23$\pm$0.03 & 0.39$\pm$0.04 & 0.21$\pm$0.03 & 0.40$\pm$0.05 & 0.13 & 0.08 & 0.22\\
19.5-20.0 & 0.32$\pm$0.03 & 0.40$\pm$0.04 & 0.28$\pm$0.03 & 0.27$\pm$0.03 & 0.43$\pm$0.04 & 0.30$\pm$0.03 & 0.09 & 0.18 & 0.14\\
20.0-20.5 & 0.26$\pm$0.02 & 0.43$\pm$0.03 & 0.31$\pm$0.03 & 0.18$\pm$0.02 & 0.53$\pm$0.04 & 0.29$\pm$0.03 & 0.06 & 0.21 & 0.11\\
20.5-21.0 & 0.26$\pm$0.03 & 0.44$\pm$0.04 & 0.30$\pm$0.03 & 0.21$\pm$0.03 & 0.47$\pm$0.04 & 0.32$\pm$0.03 & 0.07 & 0.17 & 0.12\\
21.0-21.5 & 0.40$\pm$0.04 & 0.32$\pm$0.03 & 0.28$\pm$0.03 & 0.30$\pm$0.04 & 0.52$\pm$0.06 & 0.18$\pm$0.03 & 0.09 & 0.29 & 0.10\\
21.5-22.0 & 0.40$\pm$0.04 & 0.38$\pm$0.04 & 0.22$\pm$0.03 & 0.23$\pm$0.04 & 0.70$\pm$0.08 & 0.07$\pm$0.02 & 0.05 & 0.35 & 0.10\\
22.0-22.5 & 0.46$\pm$0.06 & 0.36$\pm$0.06 & 0.18$\pm$0.03 & 0.35$\pm$0.07 & 0.49$\pm$0.08 & 0.16$\pm$0.04 & 0.04 & 0.24 & 0.10\\
22.5-23.0 & 0.33$\pm$0.07 & 0.49$\pm$0.09 & 0.18$\pm$0.04 & 0.31$\pm$0.08 & 0.38$\pm$0.10 & 0.31$\pm$0.08 & 0.11 & 0.08 & 0.14\\
23.0-23.5 & 0.44$\pm$0.11 & 0.38$\pm$0.11 & 0.18$\pm$0.07 & 0.43$\pm$0.17 & 0.25$\pm$0.12 & 0.32$\pm$0.14 & 0.04 & 0.06 & 0.31\\
\hline
 & 0.34$\pm$0.02 & 0.40$\pm$0.02 & 0.26$\pm$0.02 & 0.26$\pm$0.03 & 0.47$\pm$0.03 & 0.27$\pm$0.03 & & & \\
\hline
 \end{tabular}
\end{table*} 

\subsection{Luminosity Function using a three-Gaussians fit}\label{Section4.2}

To assess the potential impact of intrinsic cross-contamination among the three stellar populations when evaluating their LFs, we employed an alternative method to derive the LFs of the three MSs. This method is based on the approach detailed in \citetalias{2019MNRAS.484.4046M}.

\begin{figure*}
 \centerline{\includegraphics[width=\textwidth]{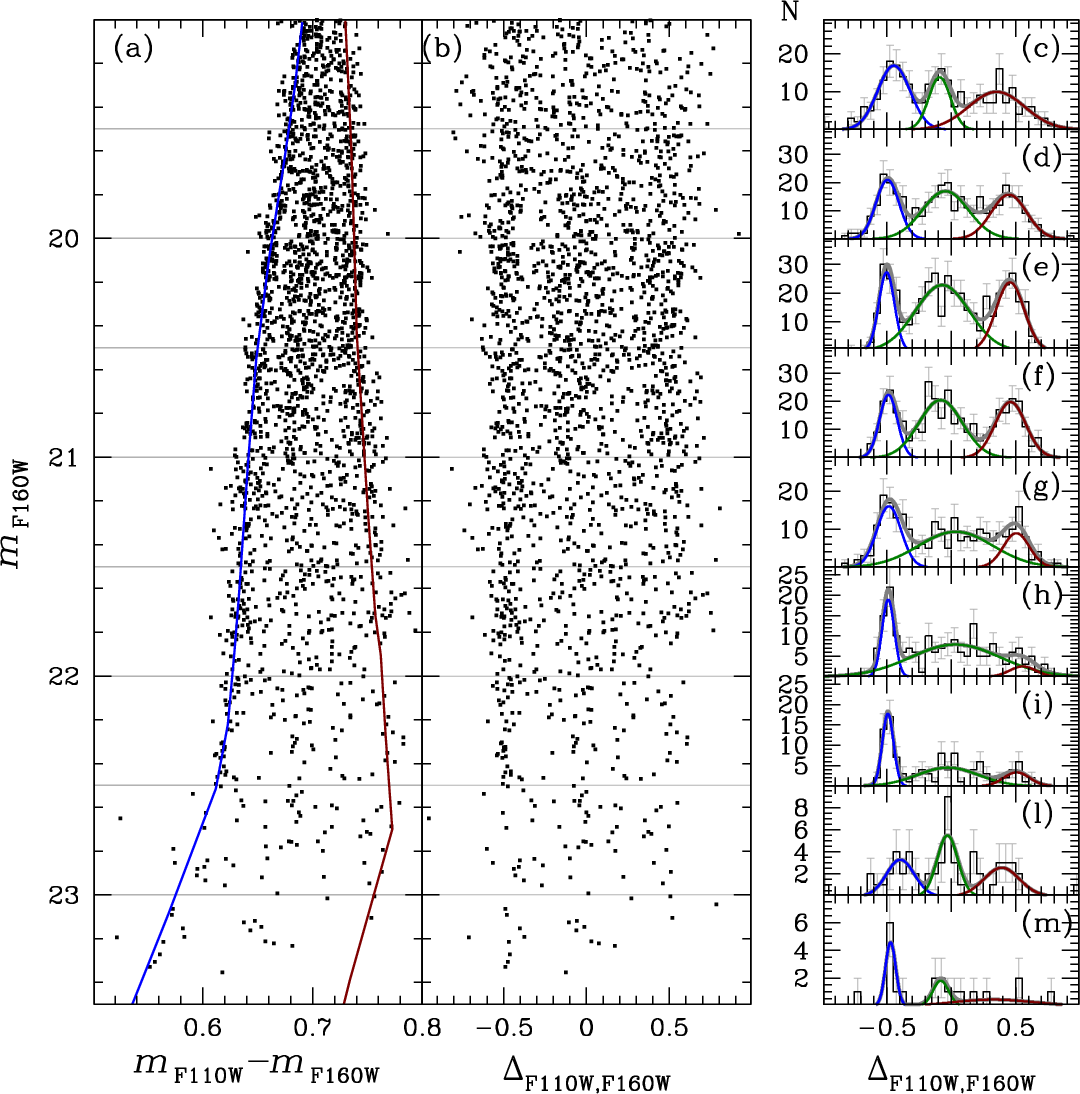}}
 \caption{(a)-(b) Same as panels\,(a) and (b) of Fig.\,\ref{Figure7}. (c)-(m) $\Delta_{\rm F110W,F160W}$ histogram distribution of cluster stars in nine luminosity intervals, as indicated by the grey lines in (a) and (b). Overlaid on each histogram are the best-fitting three-Gaussian functions, visually represented by blue, green, and red lines.} 
 \label{Figure14} 
\end{figure*}

Panels\,(a) and (b) of Fig.\,\ref{Figure14} represent the $m_{\rm F160W}$ versus $m_{\rm F110W}-m_{\rm F160W}$ CMD and the $m_{\rm F160W}$ versus $m_{\rm F110W}-m_{\rm F160W}$ verticalised CMD introduced in Fig.\,\ref{Figure7}. Panels\,(c)-(m) of Fig.\,\ref{Figure14} show the $\Delta_{\rm F110W, F160W}$ histogram distribution of stars in the nine magnitude intervals defined by the grey horizontal lines in panels\,(a) and (b) of Fig.\,\ref{Figure14}. We estimated the best-fitting three-Gaussian function for each histogram in each magnitude bin and overlaid them on the histograms as grey continuous lines, with each component denoted by blue, green, and red colours.

\begin{figure*}
 \centerline{\includegraphics[width=\textwidth]{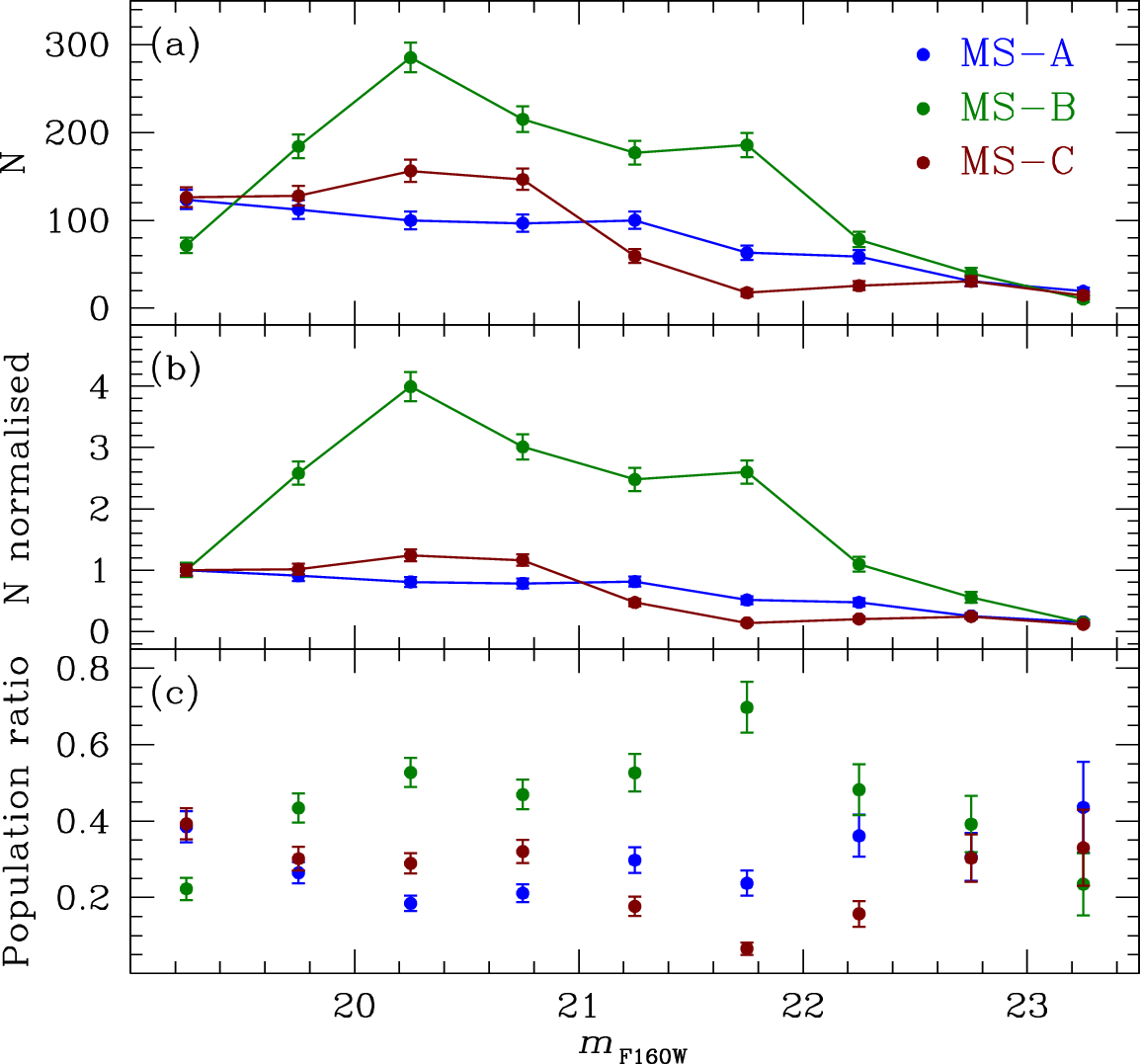}}
 \caption{Same as Fig.\,\ref{Figure9} but using a three-Gaussians fit (see Fig.\,\ref{Figure14}).} 
 \label{Figure12} 
\end{figure*}

In \citetalias{2019MNRAS.484.4046M}, the estimation of the number of stars for each population was based on the areas beneath their corresponding Gaussians. However, upon closer examination of the three components of the three-Gaussian fit shown in panels\,(c)-(m) of Fig.\,\ref{Figure14}, it becomes apparent that the Gaussian representing MS-A (in blue) provides a satisfactory fit to the histogram, while the Gaussian components of MS-B (in green) and MS-C (in red) do not adequately fit the data; for MS-C this is particularly true at fainter magnitude bins where the statistic for this component is very low. This discrepancy suggests that MS-B stars deviate from a Gaussian distribution. Consequently, a decision was made to estimate the number of stars in the MS-A and MS-C populations based on the areas beneath their respective Gaussians, and to calculate the number of stars in MS-B as the difference, in each bin, between the total number of stars and the sum of MS-A and MS-C populations. The resulting values (corrected for completeness) are represented in panel\,(a) of Fig.\,\ref{Figure12} and listed in Table\,\ref{Table4}. Notably, the LFs shown in this figure exhibit notable differences compared to those presented in Fig.\,\ref{Figure9} and estimated using the first procedure. Specifically, there is a reduction in the number of stars in MS-A, and an increase in the number of stars in MS-B, while the number of MS-C stars has, on average, remained relatively constant. Within the magnitude interval $19 < m_{\rm F160W} < 21.25$, MS-B is the most populous, whereas MS-A and MS-C have approximately the same number of stars (albeit lower compared to MS-B).

The shape of the three LFs in Fig.\,\ref{Figure9} differs significantly, as highlighted in panel\,(b) of Fig.\,\ref{Figure12}, where the LFs for individual MS populations are normalized by the value of the first bin for each respective MS. Specifically, the LF of MS-C now appears to resemble the LF of MS-A, in contrast to the results presented in Fig.\,\ref{Figure9} where it exhibited a trend similar to MS-B. We explain this change as due to the high degree of overlap between MS-B and MS-C distributions (as shown in panels\,(c)-(m) of Fig.\,\ref{Figure14}), making it challenging to establish a precise boundary between these two populations. The observed similarity in panel\,(b) of Fig.\,\ref{Figure9} between the LFs of MS-B and MS-C may be attributed to cross-contamination of MS-B into the MS-C regions, which is particularly effective given the broad non-Gaussian distribution of MS-B and its larger population size compared to MS-C.

\begin{figure*}
 \centerline{\includegraphics[width=\textwidth]{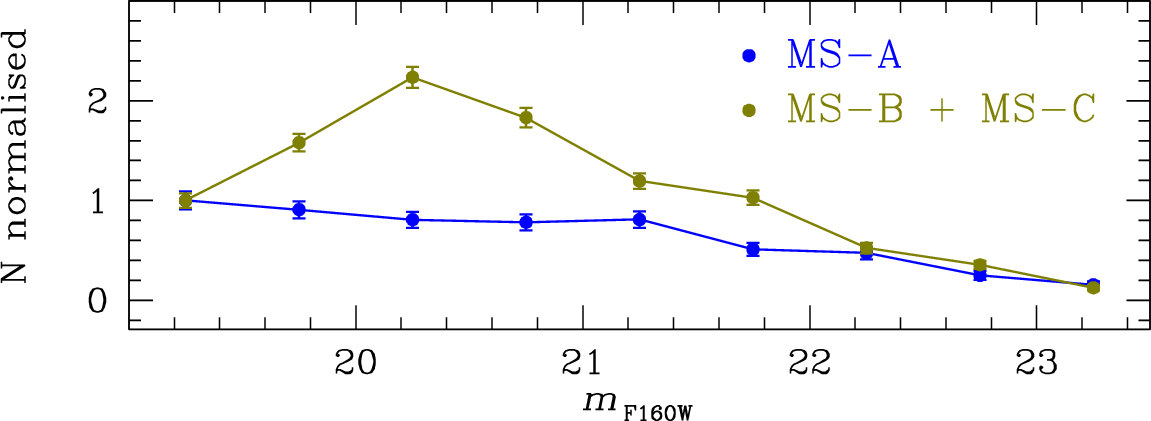}}
 \caption{LFs for the MS-A and MS-B+MS-C together, normalised by the value of the first bin for each respective LF. See text for details.} 
 \label{Figure13} 
\end{figure*}

In essence, estimating the total fraction for MS-C proved to be the most challenging task. To assess the statistical significance of the separation between the MS-B and MS-C populations we carried out a Hartigan's dip test between the two populations (for more details, please refer to our \href{https://aas243-aas.ipostersessions.com/?s=C5-BD-B8-2B-4D-32-D3-01-81-68-3C-2D-BC-25-50-1B}{poster}). The test failed with p>0.2 at the majority of stellar masses, indicating the challenge in distinguishing between the two populations. Due to the imperfect separation of the MS-B and MS-C sequences and the uncertainties involved in evaluating separate LFs for each population, we opted for a prudent approach and considered the two populations together in the LF analysis. The most cautious approach appears to be estimating the number of MS-A using a Gaussian and determining the combined number of MS-B and MS-C as the difference between the total number of stars and the MS-A population. The LF for MS-A and the combined LF for MS-B and MS-C are illustrated in Fig.\,\ref{Figure13} for each magnitude bin. The values are reported in Table \ref{Table5}. It is evident that the LF of MS-B$+$MS-C stars exhibits a distinct shape compared to the LF of MS-A stars, providing evidence of a different LF between these two groups of stars.

Panel (c) of Fig.\,\ref{Figure12} shows the fractions of MS-A, MS-B, and MS-C stars relative to the total number of MS stars ($N_{\rm A}/N$, $N_{\rm B}/N$ and $N_{\rm C}/N$). The values, along with the dispersion of each Gaussian component ($\sigma_{\rm MS-A}$, $\sigma_{\rm MS-B}$ and $\sigma_{\rm MS-C}$), are listed in Table\,\ref{Table3}. While the ratios obtained in this study align with the values reported by \citetalias{2019MNRAS.484.4046M} within the magnitude range $19 < m_{\rm F160W} < 22$, there are differences present. These inconsistencies can be ascribed to two main differences in the data set used in our work compared to \citetalias{2019MNRAS.484.4046M}. First, \citetalias{2019MNRAS.484.4046M} lacked access to proper motion data, and therefore, contamination from field objects was addressed using a statistical approach. Second, the new observations combined with the existing data were collected with different orientations and large offsets. While this has improved photometric accuracy, it has simultaneously led to a reduction in photometric precision, resulting in broader sequences.

By calculating weighted averages of the population ratios obtained in the nine magnitude intervals, we find that MS-A, MS-B, and MS-C encompass approximately 26$\pm$3$\%$, 47$\pm$3$\%$, and 27$\pm$3$\%$, respectively, of the total number of MS stars. Notably, these values are consistent with the values reported in \citetalias{2019MNRAS.484.4046M} (26.3$\pm$1.4\%, 46.9$\pm$1.3\% and 26.8$\pm$2.0\%, respectively), differing only by $0.09\,\sigma$, $0.03\,\sigma$ and $0.06\,\sigma$, respectively. Additionally, these ratios are in line with findings from studies on NGC\,6752 encompassing a wider range of masses ($0.8-0.6$ M$_{\odot}$) and various radial distances \citep{2013ApJ...767..120M,2015A&A...573A..70N}.
 
\begin{table}
\caption{Number of stars in MS-A ($N_{\rm A}$) and MS-B$+$MS-C together ($N_{\rm B}+N_{\rm C}$).}
\centering
 \label{Table5}
\begin{tabular}{c @{\quad \vline \quad} c c}
\hline
 $\Delta m_{\rm F160W}$ & & \\
\hline
  & $N_{\rm A}$ & $N_{\rm B}+N_{\rm C}$\\
 \hline
19.0-19.5 & 124$\pm$11 & 197$\pm$14\\
19.5-20.0 & 112$\pm$11 & 312$\pm$18\\
20.0-20.5 & 100$\pm$10 & 441$\pm$21\\
20.5-21.0 & 97$\pm$10 & 361$\pm$19\\
21.0-21.5 & 100$\pm$10 & 236$\pm$15\\
21.5-22.0 & 63$\pm$8 & 203$\pm$14\\
22.0-22.5 & 58$\pm$8 & 104$\pm$10\\
22.5-23.0 & 31$\pm$6 & 70$\pm$8\\
23.0-23.5 & 19$\pm$4 & 25$\pm$5\\
\hline
 \end{tabular}
\end{table}

\section{Mass Function of sub-populations}\label{Section5}

\begin{figure}
\centering
 \includegraphics[width=\columnwidth]{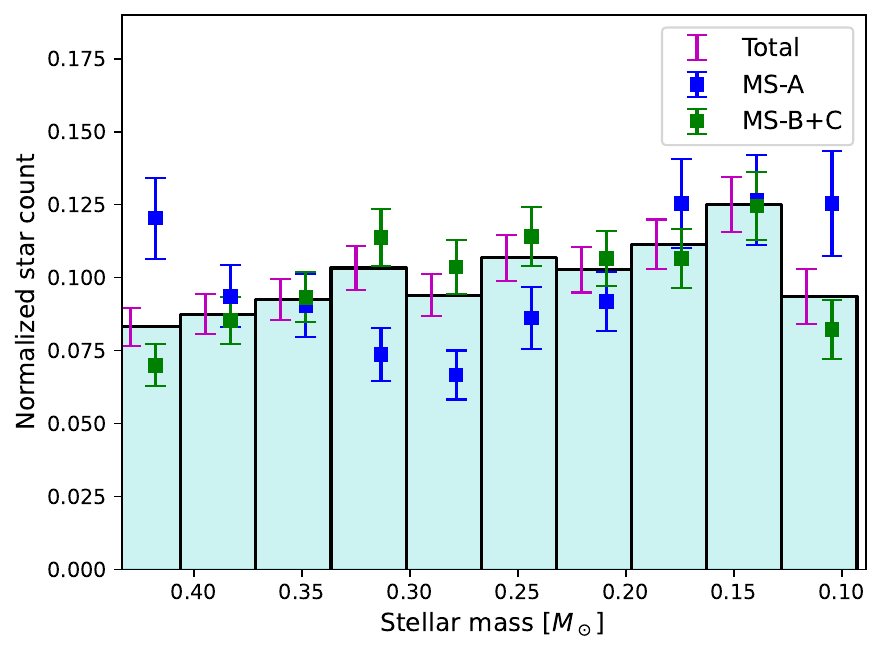}
 \caption{Estimated total mass function of NGC\,6752 (cyan histogram bins with magenta error bars), sub-population MS-A (blue markers) and combination of the other two populations (green markers). The values are normalised such that the total star count is unity for all shown MFs. Each marker refers to the star count in the entire bin; small horizontal offsets in marker positions within the bins were added for clarity.} 
 \label{fig:MF} 
\end{figure}

To determine the MF of NGC\,6752 and its populations, we estimated the mass of each star below the MS knee by identifying the closest isomass line to the star's position on the CMD. The isomass lines for given initial stellar masses were determined by fitting parabolas to the isochrone points, corresponding to those masses. The obtained mass distribution was divided into $10$ uniform bins, spanning the mass range from $\approx$ 0.091\,M$_\odot$ to 0.44\,M$_\odot$. The mass range was chosen to be as wide as possible while remaining within the magnitude bounds of Table \ref{Table5} ($19<m_{\rm F160W}<23.5$). The resulting histogram is shown in Fig.\,\ref{fig:MF}.

The star counts in each bin were corrected for the completeness estimates from Fig.\,\ref{Figure6}. The displayed errors in each bin include the Poisson counting error and the statistical error due to incomplete sampling. The latter was calculated by drawing random magnitude measurements in each mass bin according to the observed magnitude distribution and carrying out synthetic observations, using the estimated completeness in Fig.\,\ref{Figure6} as the probability for each synthetic star to be observed. The final incomplete sampling error was taken as the standard deviation of the number of observed synthetic stars over $10^5$ Monte Carlo trials.

The individual MF of MS-A and the combined MF of MS-B and MS-C were calculated by multiplying the total star count in each mass bin by the average population fraction within that bin, according to Table\,\ref{Table5}. The displayed errors for these two MFs in Fig.\,\ref{fig:MF} include the statistical error in the total star count (Poisson and incomplete sampling), as well as the error in the estimated population fraction, all added in quadrature.

To determine whether the inferred MFs of MS-A and MS-B+C are statistically consistent, we evaluated the error-scaled absolute differences between the normalized star counts in each mass bin:

\begin{equation}
    \Delta_i=\frac{\left| \Xi_i^A - \Xi_i^{B+C} \right|}{\sqrt{\mathrm{Var}(\Xi_i^A) + \mathrm{Var}(\Xi_i^{B+C})}}
    \label{MS_diff}
\end{equation}

\noindent where $\Xi_i^A$ is the completeness-corrected normalized star count of MS-A in the $i$-th mass bin, $\Xi_i^{B+C}$ is the completeness-corrected normalized combined star count of MS-B and MS-C in the $i$-th bin, $\mathrm{Var}$ represents the squared error of a given quantity. The normalization was chosen as $\sum_i \Xi_i^A=\sum_i \Xi_i^{B+C}=1$. If the MFs of MS-A and MS-B+C are consistent, $\Delta_i$ is expected to approximately follow the standard half-normal distribution. We carried out a one-sample Kolmogorov-Smirnov test between the calculated values of $\Delta_i$ and the standard half-normal distribution. The resulting $p$-value of $0.027$ suggests that the discrepancies between the MF of MS-A and the combined MF of MS-B+C cannot be explained by the measurement errors alone.

More in general, Fig.\,\ref{fig:MF} shows that the MF of MS-A tends to be flatter than that of MS-B+C in most of the mass range ($0.2 - 0.45\,{\rm M}_{\odot}$).

\section{Summary and conclusions}\label{Section6}

In this paper, we exploited NIR images of an external region of the GC NGC\,6752, which were acquired by \textit{HST} using the F110W and F160W filters of the WFC3/NIR camera at three distinct epochs (2018, 2019, and 2021). These filters are known for their sensitivity to stars with varying oxygen enhancement, making them ideal for investigating and differentiating mPOPs in the lower part of the MS.

The first epoch of this dataset was previously examined in an earlier publication within this series \citepalias{2019MNRAS.484.4046M}. In that study, it was revealed that the three distinct stellar populations within NGC\,6752, originally observed in the brighter section of the CMD, exhibit three well-defined sequences extending from the MS knee down to $\sim$0.1\,M$_{\odot}$.

In this work, we presented a more detailed investigation of these observed multiple stellar sequences by incorporating additional observational epochs. The inclusion of new observational data has enabled us to make accurate distinctions between cluster members and background/foreground sources by using PM measurements. Furthermore, the number of available images has doubled compared to the previous work in \citetalias{2019MNRAS.484.4046M}. This expanded photometric dataset enables us to conduct a more precise and comprehensive analysis of the three distinct stellar populations.

We presented the $m_{\rm F160W}$ versus $m_{\rm F110W}-m_{\rm F160W}$ CMD of the catalogue analysed in this study, which shows the triple sequence of stars extending from the MS knee to the lower end of the MS ($m_{\rm F160W}\sim24$). We utilised this observed CMD to construct and fit three isochrones corresponding to the three distinct populations, employing the methodology outlined in \citet{roman_theory_paper}. The isochrones are based on newly developed evolutionary models created with the \texttt{MESA} code and new model atmospheres calculated using \texttt{PHOENIX 15} and \texttt{BasicATLAS / ATLAS 9}. 

We also introduced a representative 12.5\,Gyr WD isochrone, computed using WD models developed by \citet{bastiwd}, which are specifically tailored for metal-poor progenitors. In the upper part of the WD isochrone, we identified three WDs that survived all the selection criteria. We utilised the WD isochrone to determine the $T_{\rm eff}$ and $g$ values for the three surviving WDs.

Our research primarily focused on investigating the LFs and MFs in the lower part of the MS for the three stellar populations within NGC\,6752. We utilised two methods to estimate the LFs of the three stellar populations. The first method involved the procedure introduced in \citet{2012A&A...537A..77M}. In the interval $19 < m_{\rm F160W} < 21.25$, MS-B has the most stars, with MS-A and MS-C having slightly fewer. In the interval $21.25 < m_{\rm F160W} < 23.5$, MS-B and MS-A have nearly the same number of stars, while MS-C has fewer. Within the magnitude range $19 < m_{\rm F160W} < 21.25,$ the three populations show distinct trends, with MS-B and MS-C increasing significantly compared to MS-A. In the magnitude range $21.25 < m_{\rm F160W} < 23.5$, the three populations exhibit a more similar trend.

The second approach relies on the methodology outlined in \citetalias{2019MNRAS.484.4046M}. The LFs obtained through this method exhibit significant disparities compared to those derived from the first method. Specifically, the LF of MS-C appears to resemble the LF of MS-B. This discrepancy is likely attributed to contamination, as the non-Gaussian distribution of MS-B stars may influence the observed LF of MS-C. Consequently, a decision was made to collectively consider MS-B and MS-C. The resulting LFs reveal a distinct shape for MS-B+MS-C in comparison to MS-A, indicating a difference in the LF between 1P and 2P stars.

To determine the MF of NGC\,6752 and for its populations, the mass of each star below the MS knee was estimated by identifying the closest isomass line to its position on the CMD. The mass distribution was divided into 10 uniform bins, spanning a mass range from approximately 0.087\,M$_\odot$ to 0.47\,M$_\odot$. The individual MF of MS-A and the combined MF of MS-B and MS-C were calculated by multiplying the total star count in each mass bin by the average population fraction within that bin. Our analysis revealed that the MF of MS-A differs from the combined MF of MS-B+C. In particular for masses $0.2 - 0.45\,{\rm M}_{\odot}$ the MF of MS-A tends to be flatter than the MFs of MS-B+C
%

The present-day MF of stars in a globular cluster is determined by a combination of internal and external dynamical effects: two-body relaxations drive the segregation of massive stars towards the central regions and the outward migration of low-mass stars leading to a radial variation of the local MF with the distance from the cluster centre. At the same time, stellar escape due to the effects of two-body relaxation and the external tidal field leads to the preferential loss of low-mass stars and the gradual flattening of the global MF (see e.g. \citealt{1997MNRAS.289..898V}).

A comprehensive analysis of the MF of NGC\,6752 and its link with the cluster's dynamical history would require a complete radial coverage enabling the study of both the effects of mass segregation and those associated with stellar escape \citep[see e.g.][]{2016MNRAS.463.2383W}. Our analysis is limited to the cluster's outer regions; the radial range covered by our data extends from about $2\,r_{\rm h}$ to about $3.4\,r_{\rm h}$ but most of the stars are between $2.5\,r_{\rm h}$ and $3\,r_{\rm h}$. Hence, a comprehensive investigation is beyond the scope of this paper.

The presence of multiple populations in globular clusters differing not only in their chemical abundances but also in their dynamical properties has added another layer of complexity to the investigation of the MF and its link with the cluster's dynamical evolution.
In particular, a number of theoretical studies of the formation of multiple populations (see e.g. \citealt{2008MNRAS.391..825D,2010ApJ...724L..99B,2011MNRAS.412.2241B,2019MNRAS.489.3269C,2022MNRAS.517.1171L}) predict that 2P stars formed more centrally concentrated than 1P stars and may be characterized by kinematic differences which are either imprinted at the time of the cluster's formation (see e.g. \citealt{2010ApJ...724L..99B,2011MNRAS.412.2241B,2015MNRAS.450.1164H,2022MNRAS.517.1171L}) and/or produced during the cluster's evolution (see e.g. \citealt{2019MNRAS.487.5535T,2021MNRAS.502.4290V,2021MNRAS.502.1974S}).

A first investigation of the implications of the complex dynamical properties of multiple-population clusters for the evolution of the 1P and 2P MFs has been presented by \citet{2018MNRAS.476.2731V}. Assuming the two populations formed with the same initial MF, even in cases when the global MF does not evolve significantly, significant differences may develop between the local MF of the 1P and the 2P. In particular, as shown in the N-body simulations of \citet{2018MNRAS.476.2731V}, in a cluster's outer regions the local MF of the 2P tends to be steeper than the 1P MF. This is the consequence of the effects of mass segregation/low-mass star outward migration in a system in which the 2P was initially more centrally concentrated and populated the outer regions preferentially with low-mass stars (see \citealt{2018MNRAS.476.2731V}).

Although simulations specifically tailored to model the evolution of NGC\,6752 would be required for a detailed comparison between simulations and observations, the general trend of the difference between the 1P and the 2P MFs in the cluster's outer regions found in \citet{2018MNRAS.476.2731V} is consistent with that revealed in our observational data.  

Our observations thus provide possible evidence of one of the manifestations of the effects associated with the different dynamical histories of the different stellar populations.

Future observations, in particular with the \textit{JWST}, would be essential to further extend the LF and MF analyses of NGC\,6752 and its mPOPs, even into the brown dwarf regime. These observations in combination with a study of the structural and kinematic properties of multiple stellar populations will offer the information necessary to build a complete dynamical picture of this cluster and provide key constraints for theoretical studies of its formation and dynamical history.


\section*{Acknowledgments}
Michele Scalco and Luigi Rolly Bedin acknowledge support by MIUR under the PRIN-2017 programme \#2017Z2HSMF, and by INAF under the PRIN-2019 programme \#10-Bedin. EV acknowledges support from NSF grant AST-2009193. R.\ Gerasimov and A. Burgasser acknowledge funding support from Hubble HST Programs GO-15096 and GO-15941, provided by NASA through a grant from the Space Telescope Science Institute, which is operated by the Association of Universities for Research in Astronomy, Incorporated, under NASA contract NAS5-26555. A. Bellini acknowledges support from HST programs GO-15096 and GO-15857.

\section*{Data Availability}
The data underlying this article were accessed from the Mikulski Archive for Space Telescopes (MAST), available at \url{https://archive.stsci.edu/hst/search.php}. All data come from the HST programme GO-15096 + GO-15491 (P.I.:\,Bedin). The full list of observations is reported in Table\,\ref{Table1}.

\bibliography{Wiley-ASNA}%

\end{document}